\documentclass[amsmath,amssymb]{revtex4}

\usepackage{color}

\usepackage{graphicx}
\usepackage{dcolumn}
\usepackage{bm}
\input epsf

\def\u{{\rm \bf u}}

\newcommand{\ksnew}{\textcolor{black}}
\usepackage{color} 

\begin{document}
\title{Shapes and paths of an air bubble rising in quiescent liquids}
\author{D. M. Sharaf$^{\dagger\dagger}$, A. R. Premlata, Manoj Kumar Tripathi$^{\dagger}$, Badarinath Karri$^{\dagger\dagger}$ and Kirti Chandra Sahu\footnote{ksahu@iith.ac.in}}
\address{Department of Chemical Engineering, Indian Institute of Technology Hyderabad, Kandi, Sangareddy 502 285, Telangana, India\ \\
$^{\dagger}$Indian Institute of Science Education and Research Bhopal 462 066, Madhya Pradesh, India \\
$^{\dagger\dagger}$Department of Mechanical {and Aerospace} Engineering, Indian Institute of Technology Hyderabad, Kandi, Sangareddy 502 285, Telangana, India}

\date{\today}

\begin{abstract}
Shapes and paths of an air bubble rising inside a liquid are investigated experimentally. About three hundred experiments are conducted in order to generate a phase plot in the Galilei and E\"{o}tv\"{o}s numbers plane, which separates distinct regimes in terms of bubble behaviour. A wide range of the Galilei and E\"{o}tv\"{o}s numbers are obtained by using aqueous glycerol solutions of different concentrations as the surrounding fluid, and by varying the bubble size. The dynamics is investigated in terms of shapes, topological changes and trajectories of the bubbles. Direct numerical simulations are conducted to study the bubble dynamics, which show excellent agreement with the experiments. To the best of our knowledge, this is the first time an experimentally obtained phase plot showing the distinct behaviour of an air bubble rising in a quiescent medium is reported for such a large range of Galilei and E\"{o}tv\"{o}s numbers. 
\end{abstract}

\maketitle


\section{Introduction}
\label{sec:intro}
The dynamics of an air bubble rising in a liquid has been an active area of research due to its relevance in many natural and industrial applications (see for instance \cite{hadamard1911,bhaga1981,magnaudet2007,tripathiNcomms2015}). In dimensionless formulation, a rising bubble can be completely described by four dimensionless numbers: the Galilei number $\left(Ga (\equiv \rho_o g^{1/2} R^{3/2}/\mu_o)\right)$, E\"{o}tv\"{o}s number, $\left(Eo (\equiv \rho_o g R^2/\sigma) \right)$, density ratio $\left(\rho_r (\equiv \rho_i/\rho_o )\right)$ and viscosity ratio $\left(\mu_r (\equiv \mu_i/\mu_o )\right)$. Here, $R$ is the equivalent radius of the bubble, $\sigma$ is the interfacial tension, while $\rho_o$, $\mu_o$, and $\rho_i$, $\mu_i$ are densities and viscosities of the continuous and dispersed phases, respectively. An additional dimensionless parameter, Morton number ($Mo$) can also be defined as $Eo^3/Ga^4 (\equiv g \mu_0^4/\rho_o \sigma^3)$, which is unique for a particular fluid as it depends on fluid properties alone.

\begin{figure}[h]
\centering
 \includegraphics[width=0.4\textwidth]{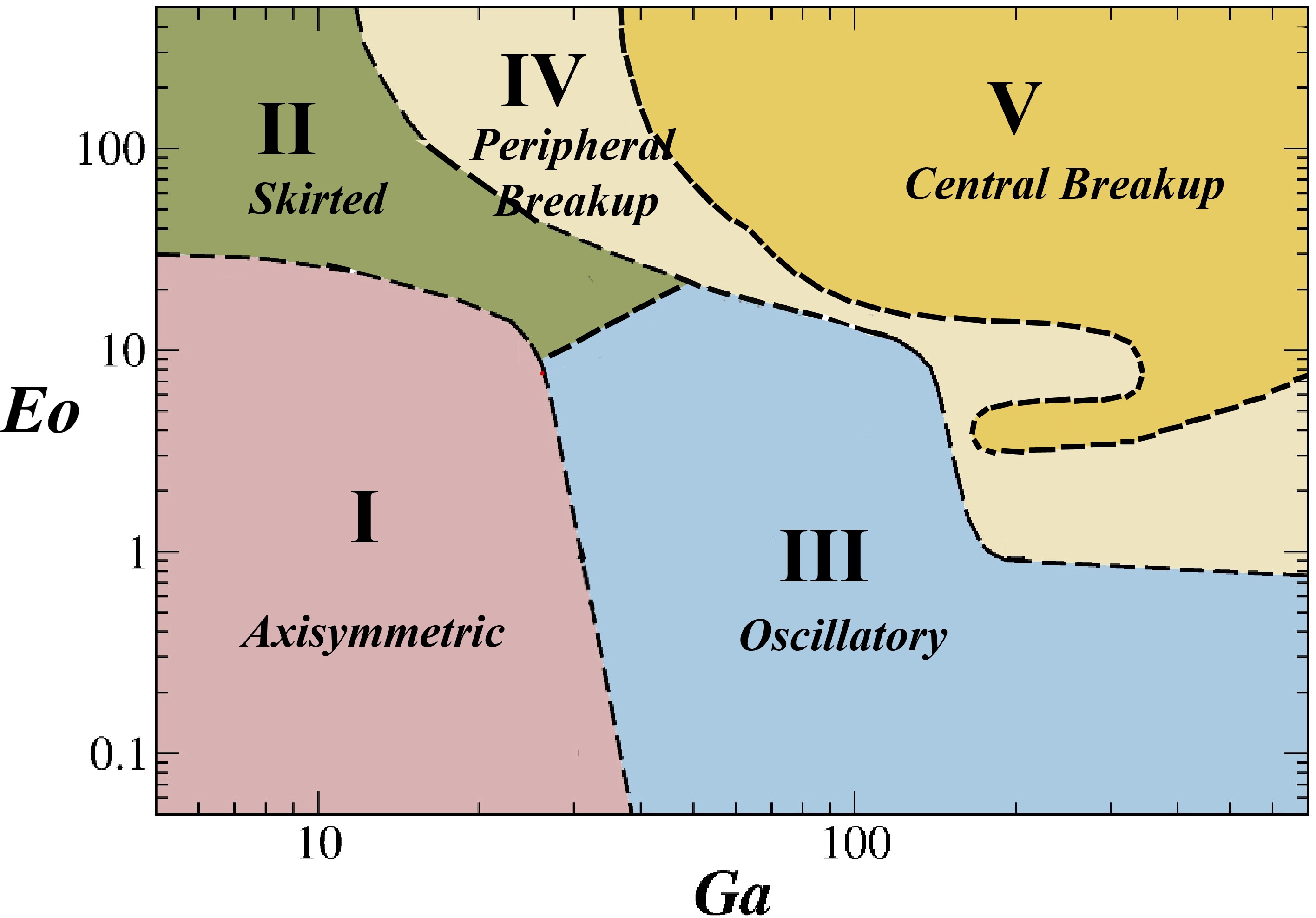}
\caption{Modified phase plot of Tripathi {\it et al.} \cite{tripathiNcomms2015} showing different regions of bubble behaviours. }
\label{mt}
\end{figure}

Recently, Tripathi {\it et al.} \cite{tripathiNcomms2015} conducted three-dimensional numerical simulations by varying $Ga$ and $Eo$ for an air bubble rising in liquid for $\rho_r=10^{-3}$ and $\mu_r = 10^{-2}$. They identified five different regions of distinct bubble behaviours (namely, axisymmetric, skirted, zigzagging/spiralling, peripheral break-up and central breakup). A sketch of these regions is shown in Fig. \ref{mt}. They showed that an air bubble maintains its azimuthal symmetry (region I) for low $Ga$ - low $Eo$, and is either spherical, oblate or dimpled. For low $Ga$, and high $Eo$ (region II), skirted bubbles are observed, whereas for high $Ga$, and low $Eo$ (region III), a bubble follows a spiral or zigzag path (wobbling motion). An air bubble with high $Ga$ and high $Eo$ breaks to form satellite bubbles (region IV) or undergoes topological changes to form a toroidal shape (region V; central breakup). In the present work, we complement the simulation results of Tripathi {\it et al.} \cite{tripathiNcomms2015} by conducting extensive experiments over the entire range of $Ga$ and $Eo$ reported by them, and provide an experimentally obtained phase plot in $Ga$-$Eo$ plane. 

In this context, it is important to discuss the classical region map provided by Bhaga \& Weber \cite{bhaga1981} (also see Ref. \cite{clift1978}), which has been used by several researchers, although Haberman \& Morton \cite{haberman1953} were probably the first to conduct experiment on rising bubble in viscous liquids. Bhaga \& Weber \cite{bhaga1981} and Clift {\it et al.} \cite{clift1978} conducted experiments on an air bubble rising in aqueous sugar solutions of differing concentrations. They identified regimes of spherical, oblate, wobbling and skirted bubbles in the Reynolds and E\"{o}tv\"{o}s numbers plane based on the bubble shapes and motion. This phase plot was prepared using three dimensionless parameters: the Reynolds, E\"{o}tv\"{o}s and Morton numbers. The Reynolds number was defined based on terminal velocity of the bubble, whereas in the present study, we use the Galilei number, which is similar to Reynolds number, but uses $\sqrt{gR}$ instead, as the velocity scale. Consequently, our phase plot can include unsteady bubbles, for which there is no terminal velocity. The use of Galilei and E\"{o}tv\"{o}s numbers (as in our region map) gives another advantage. As an example, Landel {\it et al.} \cite{landel2008} showed that for the same volume of air (i.e. constant Galilei and E\"{o}tv\"{o}s  numbers) spherical cap bubbles with a range of rise velocities (multiple Reynolds numbers) and volume of satellite bubbles can be produced. Therefore, use of Reynolds and E\"{o}tv\"{o}s numbers (as done by Clift {\it et al.} \cite{clift1978}) may provide a multivalued nature to the phase-plot. Another consequence of this behaviour can be observed in their region map. Clift {\it et al.} \cite{clift1978} provided only approximate boundaries for unbroken bubbles (i.e., for regions I, II and III). On the other hand, our phase plot provides distinct boundaries delineating all the regions, including breakup bubbles. Thus, the phase plot presented in the present study is an useful extension to the classical region map of Bhaga \& Weber \cite{bhaga1981} or Clift {\it et al.} \cite{clift1978}.

Computational researchers (e.g. Refs. \cite{hua2008,ohta05a,Tsamopoulos2008,premlata15,premlata17,premlata17b}), who used the previous experimental results  (Refs. \cite{clift1978,bhaga1981}) to validate their numerical solvers  assumed the bubble behaviour presented by them to be true for all bubbles regardless of the mass density and viscosity ratios. The present study shows that the viscosity and/or density ratios can affect the bubble dynamics. The usage of the Reynolds and/or Weber numbers in previous studies (e.g. Refs. \cite{Tsamopoulos2008,ohta12}) poses an additional limitation for unsteady bubbles because these parameters depend on the terminal velocity, which is not easy to determine accurately in experiments and is also not known {\it a priori} in numerical simulations. Other experimental and numerical studies (e.g. Refs. \cite{ohta12,tomiyama2002,bonometti2007,dij2010a,ohta14}) focused on individual regimes of bubble behaviour by considering limited sets of parameters. Particularly, we find several papers focusing on the axisymmetric and zigzagging/spiralling regions, and/or the boundary separating these two.

In this paper, we present results from an extensive experimental study of air bubbles rising in aqueous glycerol solutions of different concentrations. A phase plot in $Ga$-$Eo$ plane is presented that shows the distinct regions based on shape and path of a rising bubble. The behaviour of an air bubble in these regions is investigated and compared with the corresponding numerical simulations. The similarities and differences between the experimental and numerical results are discussed in detail. To the best of our knowledge, none of the previous studies have shown an experimentally obtained phase plot for such a large range of Galilei and E\"{o}tv\"{o}s numbers. 

\section{Experimental set-up}
\label{sec:expt}

\begin{figure}
\centering
\includegraphics[width=0.6\textwidth]{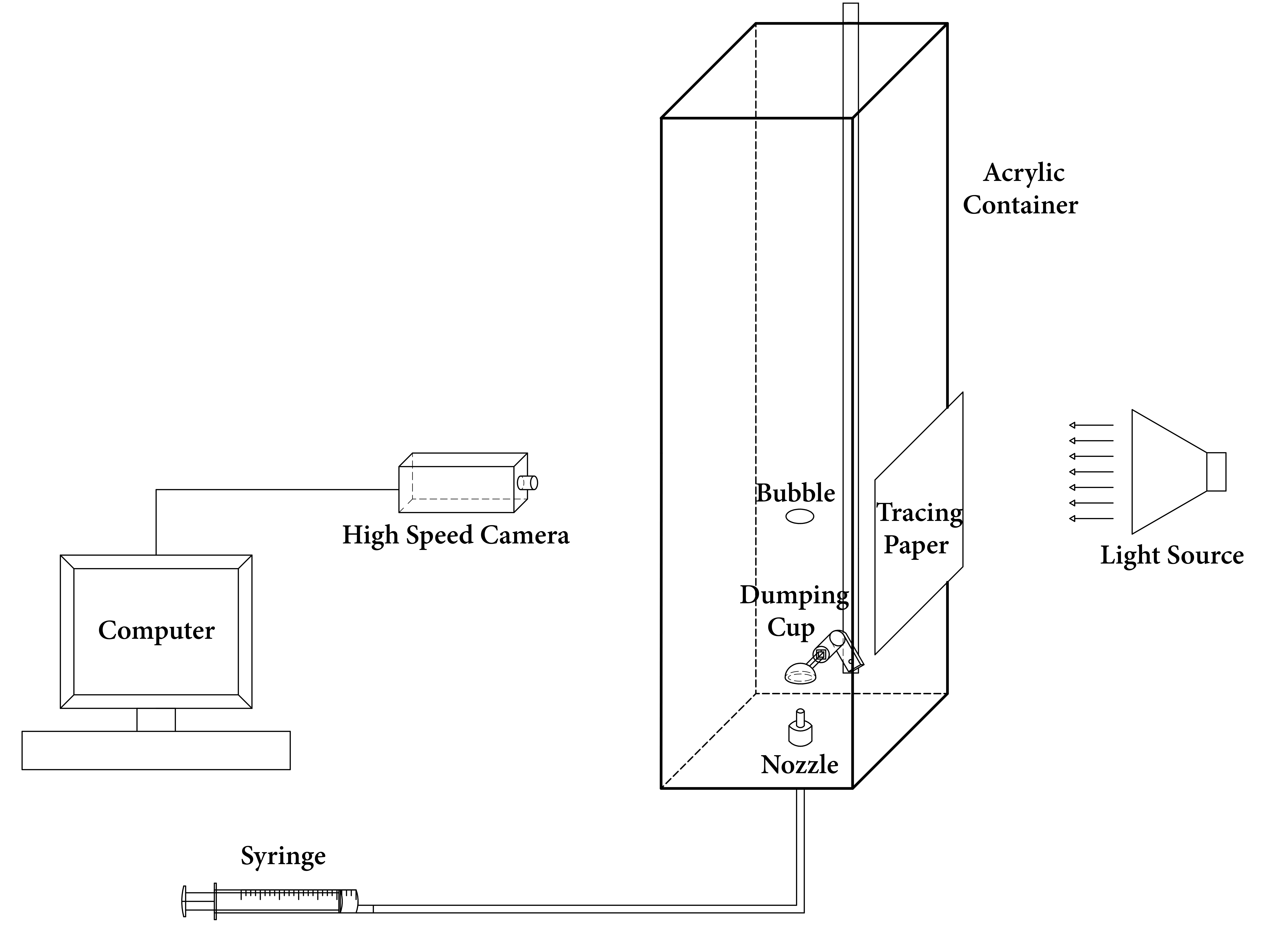} 
\caption{Schematic diagram (not to scale) showing the experimental set-up. The dimension of the acrylic tank is 200 $mm$ $\times$ 200 $mm$ $\times$ 700 $mm$.}
\label{schematic}
\end{figure}

The experimental set-up consists of (i) an acrylic tank of size 200 mm $\times$ 200 mm $\times$ 700 mm, (ii) a stainless steel nozzle (inner diameter 0.6 $mm$) fitted at the center of the bottom wall of the tank and connected to a syringe (50 $mL$ capacity) at its other end, (iii) a hemispherical dumping cup mechanism, and (iv) a high-speed camera (Photron FASTCAM SA1.1) along with back-lit illumination system and a computer. The schematic diagram of the experimental setup is shown in Fig. \ref{schematic}. The acrylic tank is used to hold an aqueous solution of glycerol with ultra pure millipore water (purity of $18.2$M$\Omega$), which acts as a surrounding fluid. A total of 19 different concentrations of glycerol in water is used to obtain a wide range of $Ga$ and $Eo$ numbers. The viscosities of these solutions are measured using a MCR 301 rheometer by Anton Paar equipped  with a cone-and-plate geometry (diameter: $40$ $mm$, angle: $0.034$ radian) at a controlled temperature of 303 $K$, whereas the other fluid properties, such as mass density and surface tension are taken from the literature (http://www.aciscience.org/docs/physical-properties-of-glycerine-and-its-solutions.pdf). The properties of these solutions are given in Table \ref{table:glycerol_solution}. The fluid in the tank is kept at a constant height of 300 mm (from the bottom of the tank) for all experiments. The cross-sectional dimensions of the tank are chosen such that the distance from the bubble to tank wall is about 10 times the maximum bubble radius used in our experiments. This minimises the wall effect on the bubble dynamics. 

A stainless steel nozzle along with the dumping cup mechanism is used to create bubbles of different sizes inside the tank. The top end of the nozzle extends to a height of about 30 $mm$ from the bottom wall of the tank at its center. The other end of the nozzle is connected to a syringe through a Poly-Tetra-Fluoro-Ethylene (PTFE) tubing. In order to create the bubbles, air is filled inside the syringe and is released by pushing the plunger with the help of a syringe pump. The air comes out through the nozzle opening inside the tank in the form of individual bubbles. These bubbles are small, spherical in shape and consistent in size ($\sim 1.4$ $mm$ radius). For creating bubbles of larger sizes, a dumping cup mechanism is used. It consists of three components: (i) a cup shaped part with a hemispherical dome in the end, (ii) a holder, which is connected to the bottom wall of the tank through a ball bearing, and (iii) a rod connected to one end of the holder for rotating the cup. The dumping cup is turned manually with the help of the rod, such that air is released and rises as a single bubble of a larger size. Note that rotating the dumping cup creates a lateral thrust on the bubble. However, we control this effect by rotating the dumping cup very slowly. It is also observed that the disturbances created by the dumping cup die down quickly as the bubble starts rising. Although the individual bubbles leaving from the nozzle are small, when they are collected inside the dumping cup they coalesce together to form a larger air bubble. The size of big bubble is calculated from the volume of air collected in the dumping cup as a volume of a sphere of an equivalent volume. The maximum equivalent radius of the bubble used in this experiment is 26.7 $mm$. 

Almost all the experimental techniques in bubble dynamics uses the differences in refractive indices of water and air to visualize the bubble. In our set-up, the bubbles are recorded by using the high-speed camera (Photron FASTCAM SA1.1), which is capable of capturing 675000 frames/second (fps) at reduced resolutions. In all our experiments, a resolution of $448 \times 800$ pixels and a frame rate of 3000 fps are used. This camera is connected through a LAN port to a computer with Photron FASTCAM Viewer (PFV) application installed in it. The camera can be controlled by using this PFV software. Two LED lighting systems along with the controller (Videoflood Controller by Visual Instrumentation Corporations) were installed opposite to the camera. These provide illumination using a diffused back-lighting method (using a tracing paper), which allows a clear visualization of the bubble boundaries on a light background.

\begin{table}
\centering
\begin{tabular}{lc c c c c c}
Sample & \% of    & $\mu_0$ & $\rho_0$ & $\sigma$  & $Mo$ \\[0.5ex]
number            & Glycerol & (mPa$\cdot$s) &  (Kg/m$^3$)& (mN/m)& \\[0.5ex]
\hline
1 &	100 &	1657 &	1260 &	62.1 & 230.314 \\
2 &	99.8 &	1524 &	1259 &	62.2 & 163.38 \\
3 &	98.2 &	1115 &	1256 &	62.3 & 46.48 \\
4 &	97 &	967.8 &	1254 &	   62.4 & 28.259\\
5 &	96 &	797 &	1249 &	      62.6 & 12.9\\
6 &	94.8 &	681 &	1246 &	  62.8 & 6.83\\
7 &	93.7 &	581 &	1243 &	63.0 & 3.6\\
8 &	92.2 &	478 &	1241 &	63.1 & 1.6\\
9 &	90.8 &	319.7 &	1235 &	63.4 & 0.3256\\
10 & 88.5 &	258  & 1230 &	63.6 & 0.1372\\
11 & 85 & 170 &	1222 &	64.2 &	0.0253\\
12 &	80 &	96.9 &	1209 &	64.8 &	0.00263\\
13 &	70 &	57.8 &	1182 &	65.8 &	0.000324\\
14 &	60 & 26 &	1154 &	66.6 &	$1.315 \times 10^{-5}$ \\
15 &	50 &	15.1 &	1127 &   67.5 &	$1.5 \times 10^{-6}$ \\
16 &	40 &	9.6 &	1100 &	68.4 &	$2.4 \times 10^{-7}$ \\
17 &	25 &	7 &	1061 &	69.5 &	$6.57 \times 10^{-8}$ \\
18 &	10 &	4.3 &	1023 &	69.8 &	$9.56 \times 10^{-9}$ \\
19 &	Pure water &	1 &	1000 &	72.8 &	$2.52 \times 10^{-11}$ \\ [1ex]
\hline
\end{tabular}
\caption{Properties of different aqueous solution of glycerol.}
\label{table:glycerol_solution}
\end{table}

\section{Numerical method}
\label{sec:num}
\begin{figure}
\begin{center}
\includegraphics[width=0.4 \textwidth]{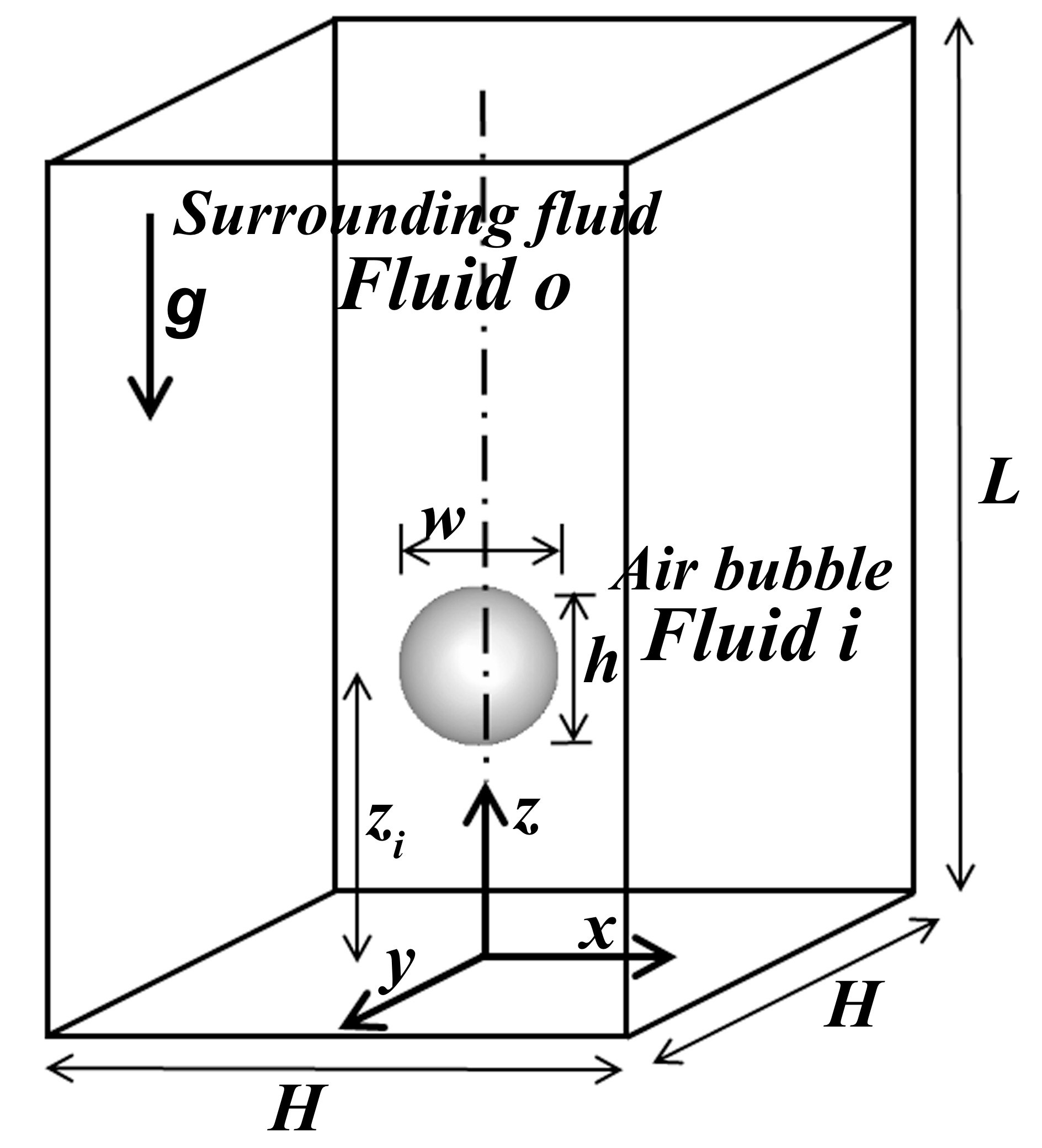}
\end{center}
\caption{Schematic diagram showing the initial configuration of an air bubble (fluid `$i$') rising in a liquid (fluid `$o$'). Initially the bubble is located at $z=z_i=15R$; $R$ being the radius of the bubble. $H = 30 R$ is the width and breadth, and $L=90R$ is the height of the rectangular computational domain. The gravity, $g$ is acting in the negative $z$ direction.}
\label{geom}
\end{figure}

Three-dimensional simulations are conducted to understand the dynamics of a rising air bubble, fluid `$i$', in a far denser and more viscous fluid, fluid `$o$', under the action of buoyancy. The schematic diagram of the computational domain is shown in Figure \ref{geom}.  An open-source fluid flow solver, Gerris created by Popinet \cite{popinet2003} is used in the present study. The present numerical solver is also the same as the one used by Tripathi {\it et al.} \cite{tripathiNcomms2015}. However, for the sake of completeness, a brief description of the numerical method is outlined below. 

A Cartesian coordinate system $(x,y,z)$ is used to model the flow dynamics. Initially, the air bubble and the surrounding fluid are stationary, with the air bubble placed at $z=z_i=15R$. Gravity acts in the negative $z$ direction. Free-slip and no-penetration conditions are imposed on all the boundaries of the computational domain. \ksnew{In our experiments, an air bubble undergoes an increase in volume as it rises. For the maximum distance travelled by a bubble in the numerical simulations (i.e. height of the computational domain), we have estimated a volume change of $<0.5$\% in the experiments. Thus, we assume the flow to be incompressible in the present numerical study. However, it is to be noted that this exercise is done only for spherical or oblate bubbles.}

The governing equations, which describe the dynamics of a rising bubble in a surrounding medium are the equations of mass and momentum conservation:
\begin{eqnarray}
\nabla \cdot \u &=& 0, \\
\rho \left [ {\partial \u \over \partial t} + \u \cdot \nabla \u \right] &=& -\nabla p + \nabla \cdot \left [\mu (\nabla \u + \nabla \u^T) \right] + \delta \sigma \kappa {\bf n} -\rho g {\bf j},
\end{eqnarray}
where ${\bf u}=(u,v,w)$ denotes the velocity field in which $u$, $v$ and $w$ represent the velocity components  in the $x$, $y$ and $z$ directions, respectively. The interface separating the air and liquid phases is obtained by solving an advection equation for the volume fraction of the liquid phase, $c$ ($c=0$ and 1 for the air and liquid phases, respectively):
\begin{equation}
{\partial c \over \partial t} + \u \cdot \nabla c  = 0,
\end{equation}
where $p$ is the pressure field, $t$ denotes time, ${\bf j}$ denotes the unit vector along the vertical direction, $\sigma$ and $g$ represent the (constant) interfacial tension for the pair of fluids considered and gravitational acceleration, respectively, $\delta$ is the Dirac delta function (given by $|\nabla c|$), whose value is one at the interface and zero otherwise. $\kappa=\nabla \cdot {\bf n}$ is the interfacial curvature, in which ${\bf n}$ is the outward-pointing unit normal to the interface. 

The mass density, $\rho$, and the dynamic viscosity, $\mu$, are assumed to depend on $c$ as
\begin{eqnarray}
\rho = c \rho_o  +  (1-c)\rho_i,\\
{\mu}=  c \mu_o + (1-c) \mu_i, 
\end{eqnarray}
where $\rho_i$, $\mu_i$ and $\rho_o$, $\mu_o$ are the density and dynamic viscosity of the dispersed (air) and the continuous (liquid) phases, respectively. 

The following scaling is used to non-dimensionalise the above governing equations:
\begin{equation}
(x,y,z) ={R} \left({\widetilde x, \widetilde y, \widetilde z}\right), \hspace{1mm} t={R \over V} \widetilde t, \hspace{1mm} {\bf u} = V\tilde{\bf u}, \hspace{1mm} p= \rho_o {V^2} \widetilde p,
\hspace{1mm} \mu = \mu_o \widetilde \mu, \hspace{1mm} \rho = \rho_{o}\widetilde \rho, \hspace{1mm} \delta = \widetilde \delta/R,
\label{eq:scaling}
\end{equation}
where the velocity scale is $V=\sqrt{gR}$, and the tildes designate dimensionless quantities. After dropping tildes from all nondimensional variables, the governing dimensionless equations are given by
\begin{eqnarray}
\nabla \cdot \u &=& 0, \\
\label{conti}
{\partial \u \over \partial t} + \u \cdot \nabla \u &=& -\nabla p + {1 \over Ga} \nabla \cdot \left [\mu (\nabla \u + \nabla \u^T) \right] + \delta {\nabla \cdot {\bf n}  \over Eo} {\bf n} - \rho {\bf j}, \label{NS} \\
{\partial c \over \partial t} + \u \cdot \nabla c &=& 0,
\label{adv1}
\end{eqnarray}
where the dimensionless density and dynamic viscosity are given by
\begin{eqnarray}
\rho = c +  (1-c) \rho_r, \\
{\mu}=  c  + (1-c) {\mu_r}.
\label{vis:model}
\end{eqnarray}

A Volume-of-Fluid (VOF) method that incorporates a height-function based balanced force continuum surface force formulation for the inclusion of the surface force term in the Navier-Stokes equation is used. In order to ensure the accuracy of the results, a dynamic adaptive grid refinement is incorporated based on the vorticity magnitude and bubble interface. This solver minimizes the amplitude of spurious currents, scaled with $\sqrt{2 R /\sigma}$, to less than $10^{-12}$. This solver was also validated extensively by comparing with the previous numerical and experimental results (see Tripathi {\it et al.} \cite{tripathiNcomms2015}).

\section{Results and Discussion}
\label{sec:dis}

\begin{figure}
\centering
 \includegraphics[width=0.6\textwidth]{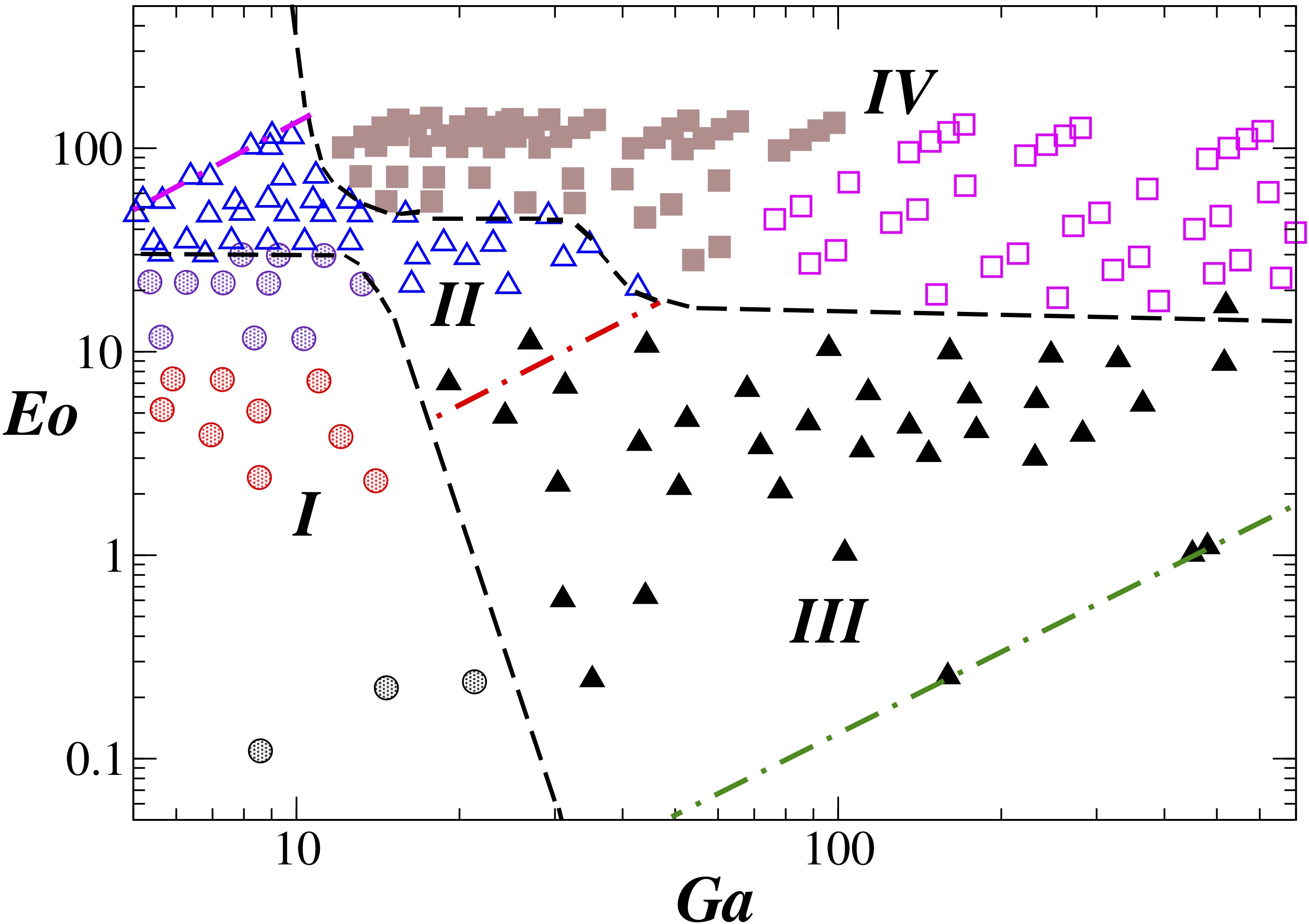}
\caption{Different regions of bubble shape and behaviour obtained from our experiments. Circles, triangles and squares represent the axisymmetric (region I), skirted (region II), oscillatory (region III) and breakup (region IV) regions, respectively. In region I, black, red and violet circles represent spherical, oblate and dimpled bubbles, respectively. In region IV, three types of breakup bubbles are observed, namely, skirted breakup (brown filled squares), satellite and toroidal breakups (magenta squares). The red dash-dotted line represents $Mo = 10^{-3}$, which separates region II and region III. The green and magenta lines represent $Mo=2.52 \times 10^{-11}$ (pure water) and $Mo=230.3$ (pure glycerol).}
\label{fig1}
\end{figure}

The dynamics of a total of 300 bubbles are analysed in terms of shapes, trajectories, break-ups and topological changes and a $Ga$-$Eo$ phase plot (Fig. \ref{fig1}) is obtained. The phase-plot was obtained experimentally by observing the rising behaviour of bubbles visually, and a minimum of 3 runs were performed for each parameter combination. Fig. \ref{fig1} shows that although regions I, II and III in the phase plot look qualitatively similar to those obtained numerically by Tripathi {\it et al.} \cite{tripathiNcomms2015} (see their Fig. 1), quantitatively there are differences. 

One major difference is that unlike five different regions of bubble behaviour in the numerical phase plot, we see only four regions in Fig. \ref{fig1}, i.e. in our experiments central breakup (region V) is not observed. The term `central breakup' has been used by the authors for the kind of breakup which results in a toroidal structure without any other significant gaseous regions forming just after the breakup. This unstable toroidal bubble further disintegrates into several smaller bubbles. This is possibly due to the difference in the initial shape of the bubble between simulations and experiments. The initial shape of the bubble used by Tripathi {\it et al.} \cite{tripathiNcomms2015} was perfectly spherical, whereas in the experiments it is not possible to inject a spherical bubble into the tank, particularly while dealing with big bubbles, i.e. for high $Ga$ and high $Eo$. The effect of initial shape of the bubble was also investigated by Ohta {\it et al.} \cite{ohta05a}. The reader may also note that none of the experimental studies so far have reported central breakup for an air bubble rising in a liquid. Another possible reason for this discrepancy could be the difference in viscosity and density ratios. In the numerical simulations of Tripathi {\it et al.} \cite{tripathiNcomms2015}), $\rho_r = 10^{-3}$ and  $\mu_r = 10^{-2}$. However, in our experiments the viscosity of the solutions used varies from $1$ mPa$\cdot$s (pure water) to 1657 mPa$\cdot$s (pure glycerol), while the density varies from 1000 kg/m$^3$ (pure water) to 1260 kg/m$^3$ (100\% glycerol). By varying the viscosity and density of the surrounding medium in our experiments, we could vary the Morton number, $Mo$ from 230.314 (for pure Glycerol) to $2.52 \times 10^{-11}$ (pure water). In addition, contamination in the tank can also alter the dynamics particularly for smaller bubbles \cite{antoine,Duineveld1995,Sanada08a}. However, we have taken care to minimise this problem by changing the liquid frequently, and minimising the time taken to carry out each experiment while allowing sufficient time for the flow to subside after pouring the liquid. All these variations contribute to the differences observed between Fig. \ref{fig1} and numerically obtained phase plot by Tripathi {\it et al.} \cite{tripathiNcomms2015}. 

In addition, we also observe some quantitative differences in terms of the actual position of the boundaries between regions. In our experimental phase plot, it is observed that region I extends upto $Ga \approx 25$ for $Eo=0.1$, but Tripathi {\it et al.} \cite{tripathiNcomms2015} observed region I upto $Ga \approx 38$ for the same value of $Eo$. Similarly for $200 \lesssim Ga \lesssim 500$ and $1 \lesssim Eo \lesssim 10$, we observe oscillatory behaviour (region III), whereas, this is reported as a break-up region in Tripathi {\it et al.} \cite{tripathiNcomms2015}.

Next, we present the bubble behaviours in each region at different dimensionless times, normalised with $\sqrt{R/g}$, and compare the dynamics with that obtained from the numerical simulations. 
\subsection{Region I bubbles}
\label{sec:region1}

\begin{figure}[h]
\begin{minipage}[b]{.3\textwidth}
(a) \\
\includegraphics[width=0.9\textwidth]{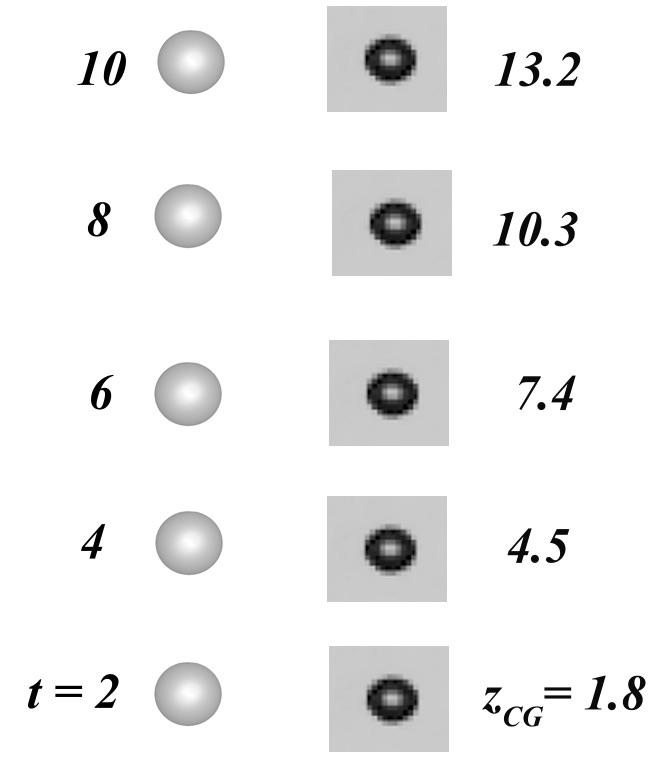}
\end{minipage} 
\begin{minipage}[b]{.3\textwidth}
(b) \\
\includegraphics[width=0.9\textwidth]{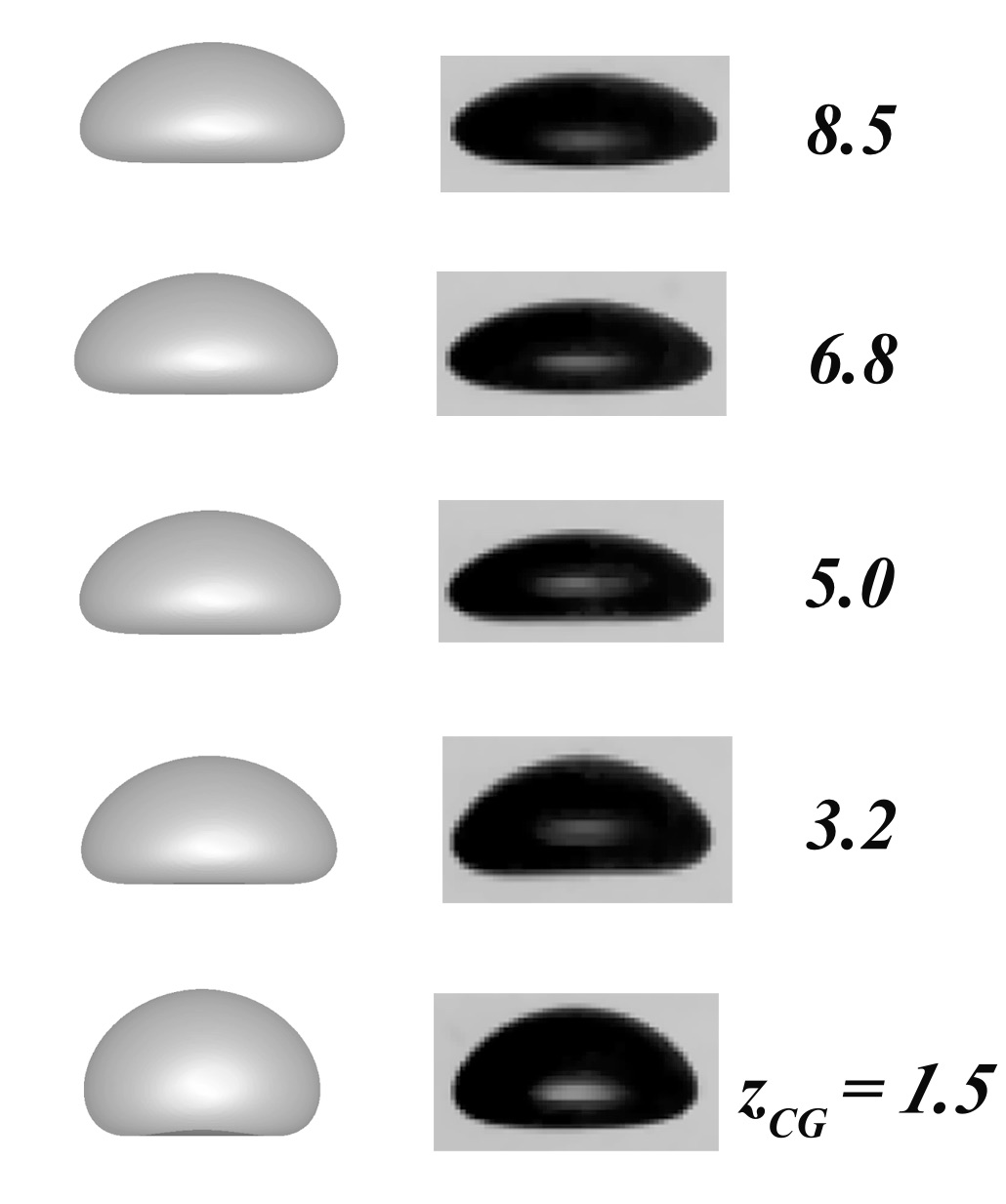} 
\end{minipage}
\begin{minipage}[b]{.3\textwidth}
(c) \\
\includegraphics[width=0.9\textwidth]{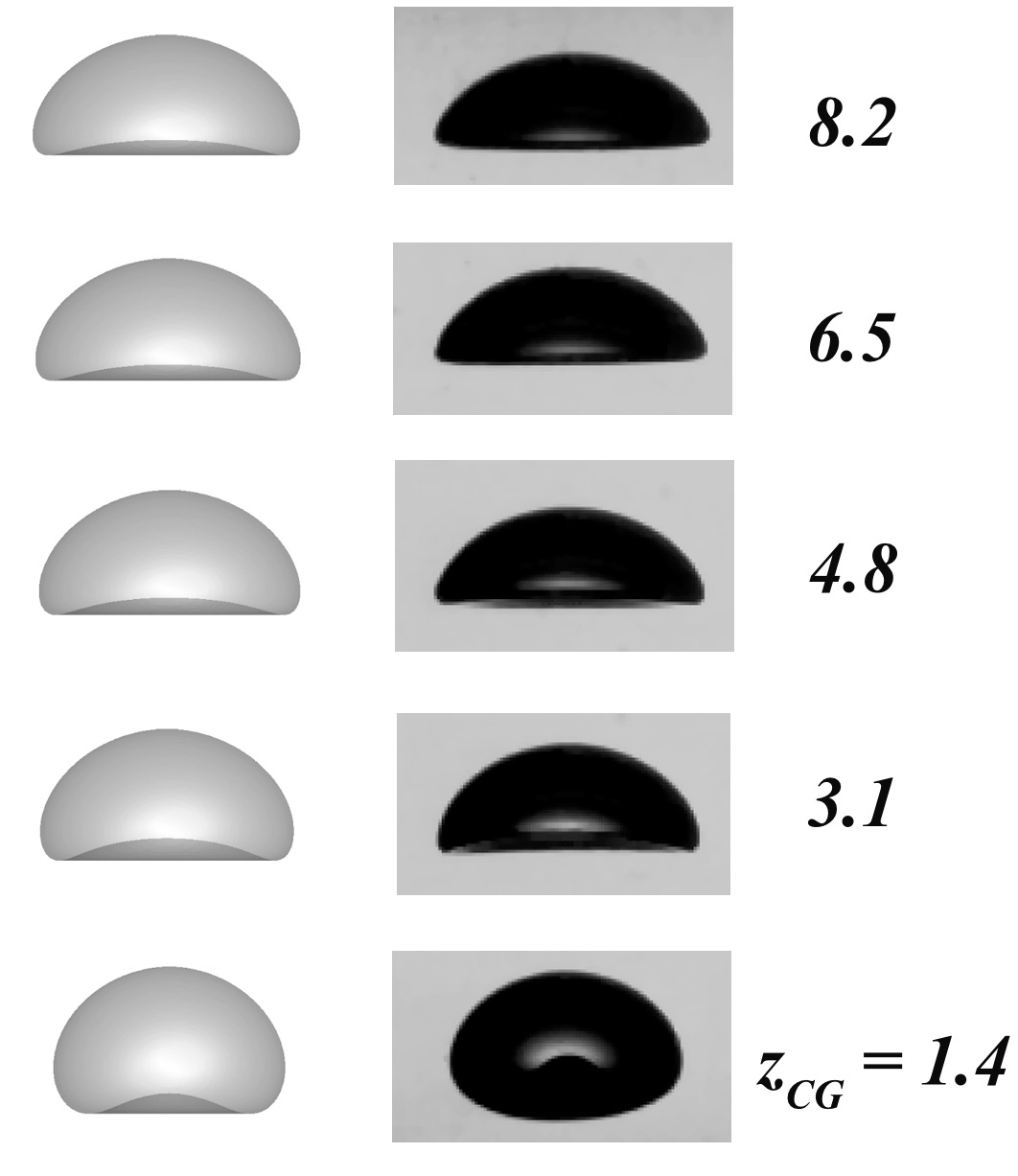} 
\end{minipage}
\caption{Evolutions of region I bubbles: (a) spherical, (b) oblate and (c) dimple shaped bubbles obtained from numerical simulations (first column in each panel) and experiment (second column in each panel). The surrounding fluid in (a), (b) and (c) are $40 \%$ of glycerol in water (i.e $\mu_r=1.04\times10^{-3} $, $\rho_r=9.09\times10^{-4}$, $Ga = 8.58$ and $Eo = 0.109$), $85 \%$ of glycerol in water (i.e $\mu_r=5.9\times10^{-5} $, $\rho_r=8.2\times10^{-4}$, $Ga = 8.52$ and $Eo = 5.11$), and $90.8 \%$ of glycerol in water (i.e $\mu_r=3.13\times10^{-5} $, $\rho_r=8.1\times10^{-4}$, $Ga = 8.36$ and $Eo = 11.7$), respectively. The radii of the bubble in panels (a), (b) and (c) are 0.83 $mm$, 5.2 $mm$ and 7.8 $mm$, respectively.}
\label{region1_bub}
\end{figure}

\begin{figure}
\centering
\includegraphics[width=0.5\textwidth]{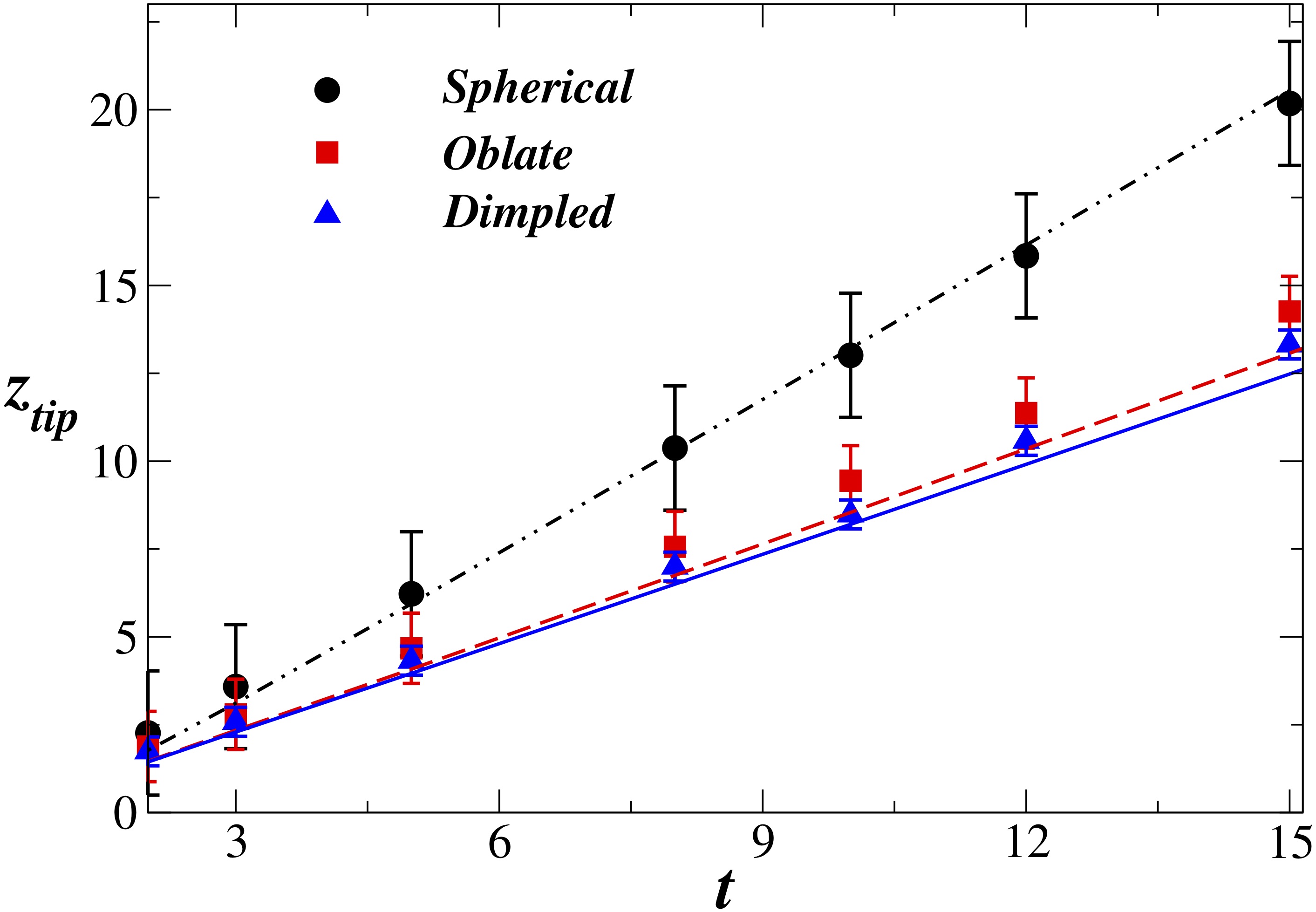} 
\caption{Temporal variations of normalised bubble-tip position, $z_{tip}$ obtained experimentally (symbols) and numerically (lines). The surrounding fluids are: $40 \%$ of glycerol in water (i.e. $\mu_r=1.04\times10^{-3} $, $\rho_r=9.09\times10^{-4}$, $Ga = 8.58$ and $Eo = 0.109$) for spherical,  $85 \%$ of glycerol in water (i.e. $\mu_r=5.9\times10^{-5} $, $\rho_r=8.2\times10^{-4}$, $Ga = 8.52$ and $Eo = 5.11$) for oblate and $90.8 \%$ of glycerol in water (i.e. $\mu_r=3.13\times10^{-5} $, $\rho_r=8.1\times10^{-4}$, $Ga = 8.36$ and $Eo = 11.7$) for dimpled bubbles. The radii of the spherical, oblate and dimpled bubbles are 0.83 $mm$, 5.2 $mm$ and 7.8 $mm$, respectively.}
\label{region1}
\end{figure}

Three types of terminal bubble shapes: spherical, oblate and dimpled can be observed in region I (axisymmetric region). Temporal evolutions of typical spherical, oblate and dimple shaped bubbles obtained from numerical simulations and experiment are shown in Fig. \ref{region1_bub}. The differences in shape are evident by visual examination. We found that for low $Eo$, the bubble remains spherical, and increasing the value of $Eo$, the bubble becomes oblate. As the $Ga$ is kept constant in Fig. \ref{region1_bub}, increasing $Eo$ (which mean decreasing the effect of surface tension) promotes deformation, as expected, which in turn changes spherical to oblate and then to dimpled bubbles. Our observations mostly agree with that of Clift {\it et al.} \cite{clift1978}, who also observed spherical bubbles for $Eo$ below $0.2$ (approximately). This region obtained from our experiment qualitatively agrees with that of Tripathi {\it et al.} \cite{tripathiNcomms2015}.

Fig. \ref{region1} shows the temporal variation of bubble-tip position normalised with the equivalent radius of the bubble ($z_{tip}$) obtained from the experiments (symbols) and numerical simulations (lines). It is observed that in region I, the rise velocity of the bubble decreases with increase in $Eo$, i.e.  a spherical (dimpled) bubble has the highest (lowest) rising velocity. This is to be expected as the drag force experienced by a spherical bubble is lower than that of an oblate bubble even if the volume remains constant. The radii of the spherical, oblate and dimpled considered in Figs. \ref{region1} and \ref{region1_bub} are 0.83 $mm$, 5.2 $mm$ and 7.8 $mm$, respectively. As the drag force increases with the increase in the size of the bubble, the drag force experienced by these bubble increases in an order from spherical to oblate to dimpled bubbles. It can also be noticed in Fig. \ref{region1} that the temporal variations of $z_{tip}$ obtained from the experiments matched with those of numerical simulations within the limit of experimental error.

\subsection{Region II bubbles}
\label{sec:region2}

\begin{figure}
\centering
\hspace{-1.0cm}  (a) \hspace{6.5cm} (b) \\
\includegraphics[width=0.3\textwidth]{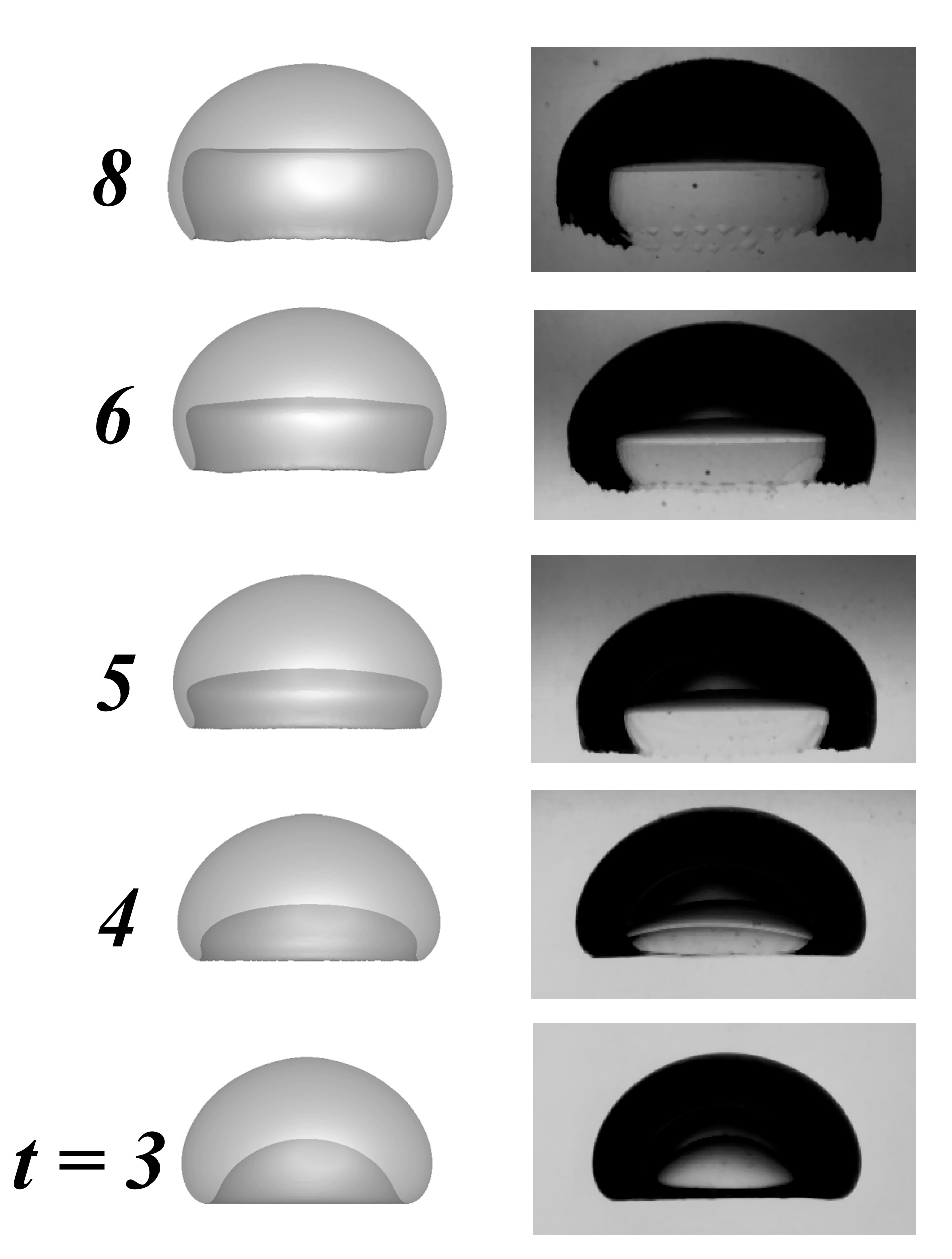} \hspace{1.0cm}
\includegraphics[width=0.5\textwidth]{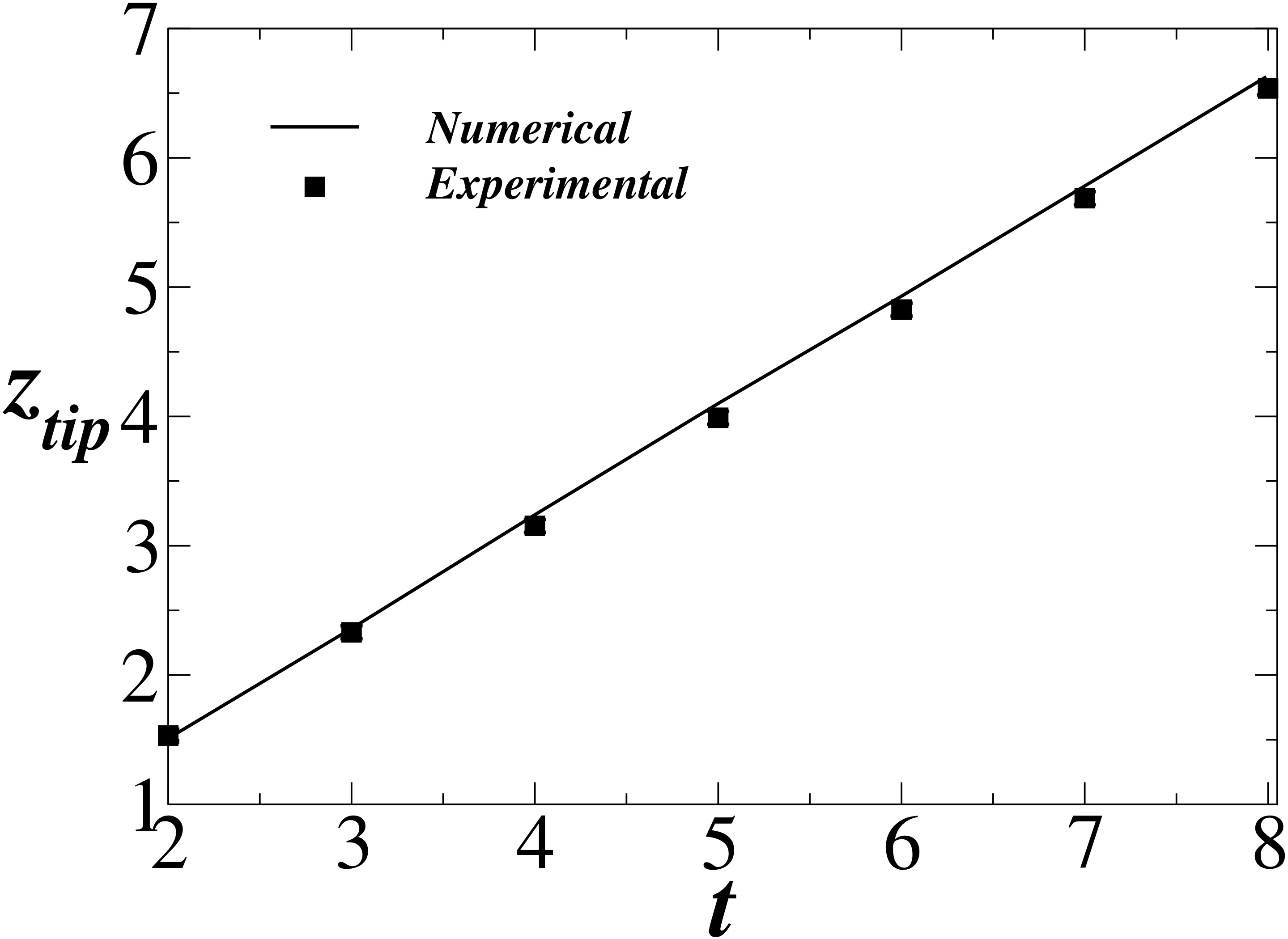} 
\caption{(a) Evolution of shapes of a typical skirted (region II) bubble. The left and right panels show the bubble shapes obtained from numerical simulation and experiment, respectively. (b) The temporal variation of $z_{tip}$ obtained from the experiment (symbols) and numerical simulation (solid line). The surrounding fluid is $97\%$ of glycerol in water ($\mu_r=1.03\times10^{-5} $, $\rho_r=7.97\times10^{-4} $, $Ga = 10.86$ and $Eo = 73.28 $). The radius of the bubble is 19.27 $mm$.}
\label{Skirt}
\end{figure}

A bubble in region II rises in a straight path, but forms a skirt-like structure along the periphery. The skirted bubbles have been investigated by a few researchers numerically (e.g. \cite{ohta12,baltussen2014,tripathiNcomms2015}) and experimentally (e.g. \cite{bhaga1981}) in the past. Fig. \ref{Skirt}(a) shows a typical region II bubble obtained from experiment (right panel) and numerical simulation (left panel). The experiment shows that as the bubble rises, a dimple forms at an early time ($t=3$). The dimple evolves into a skirt (see the translucent part of the bubble at the bottom), which grows as the bubble rises ($t\ge 4$). For this set of parameters, the skirt does not develop holes, but we see oscillations at the edge of the skirt (see $t=8$). The corresponding numerical simulation also shows very similar bubble dynamics. Both experiment and numerical simulation show an initial continuous elongation of the skirt, slight oscillation in the shape of the bubble and local oscillations at the edge of the skirt. In our experiment, we  found an excellent agreement on the time evolution of a skirted bubble (through high speed imaging) and the corresponding numerical simulation. Fig. \ref{Skirt}(b) presents the temporal variation of $z_{tip}$ obtained from the experiment and numerical simulation, which also shows a good agreement. 

\subsection{Region III bubbles}

\begin{figure}
\centering
(a) \hspace{6.0cm} (b) \\
 \includegraphics[width=0.25\textwidth]{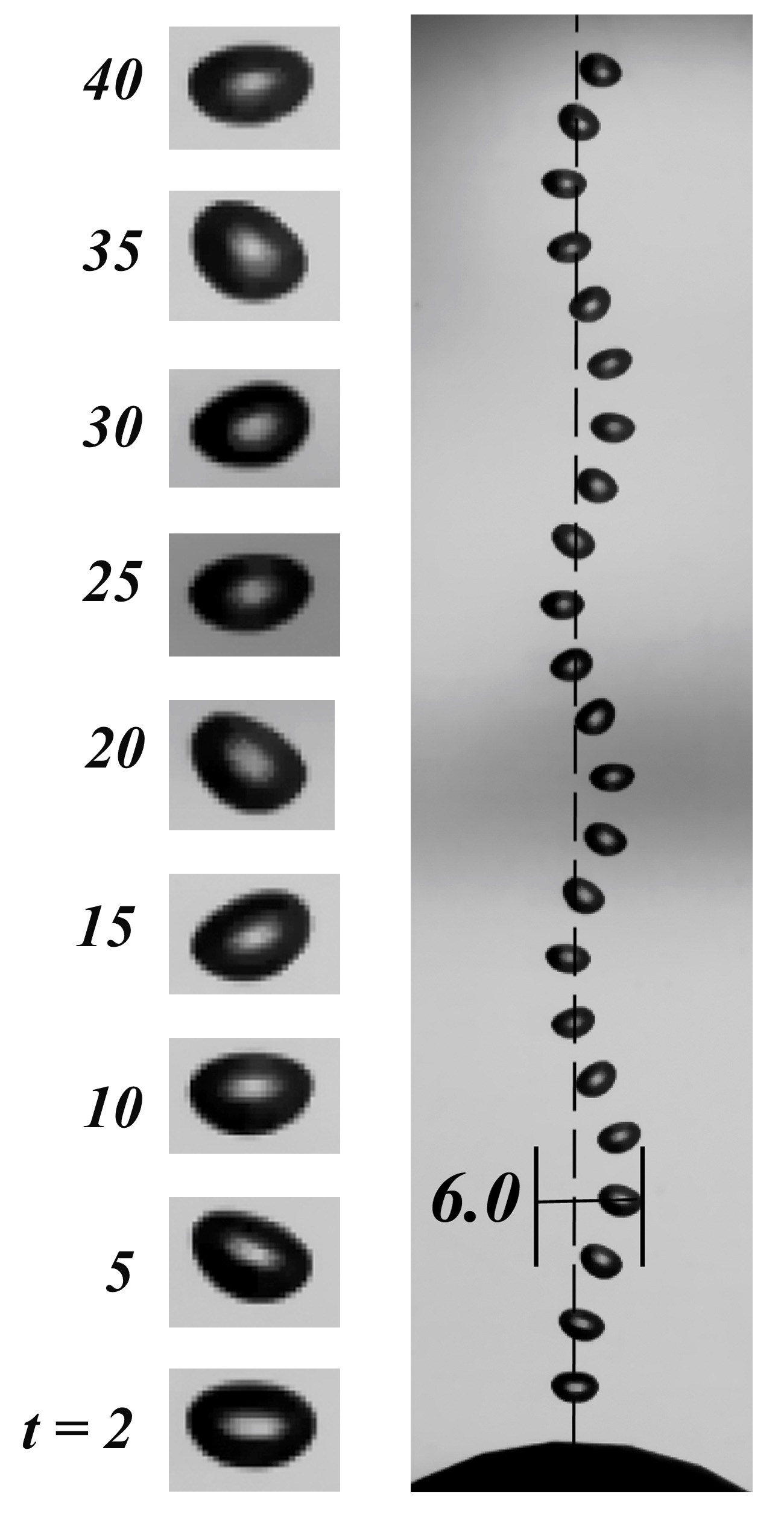}  \hspace{0.2cm}
\includegraphics[width=0.38\textwidth]{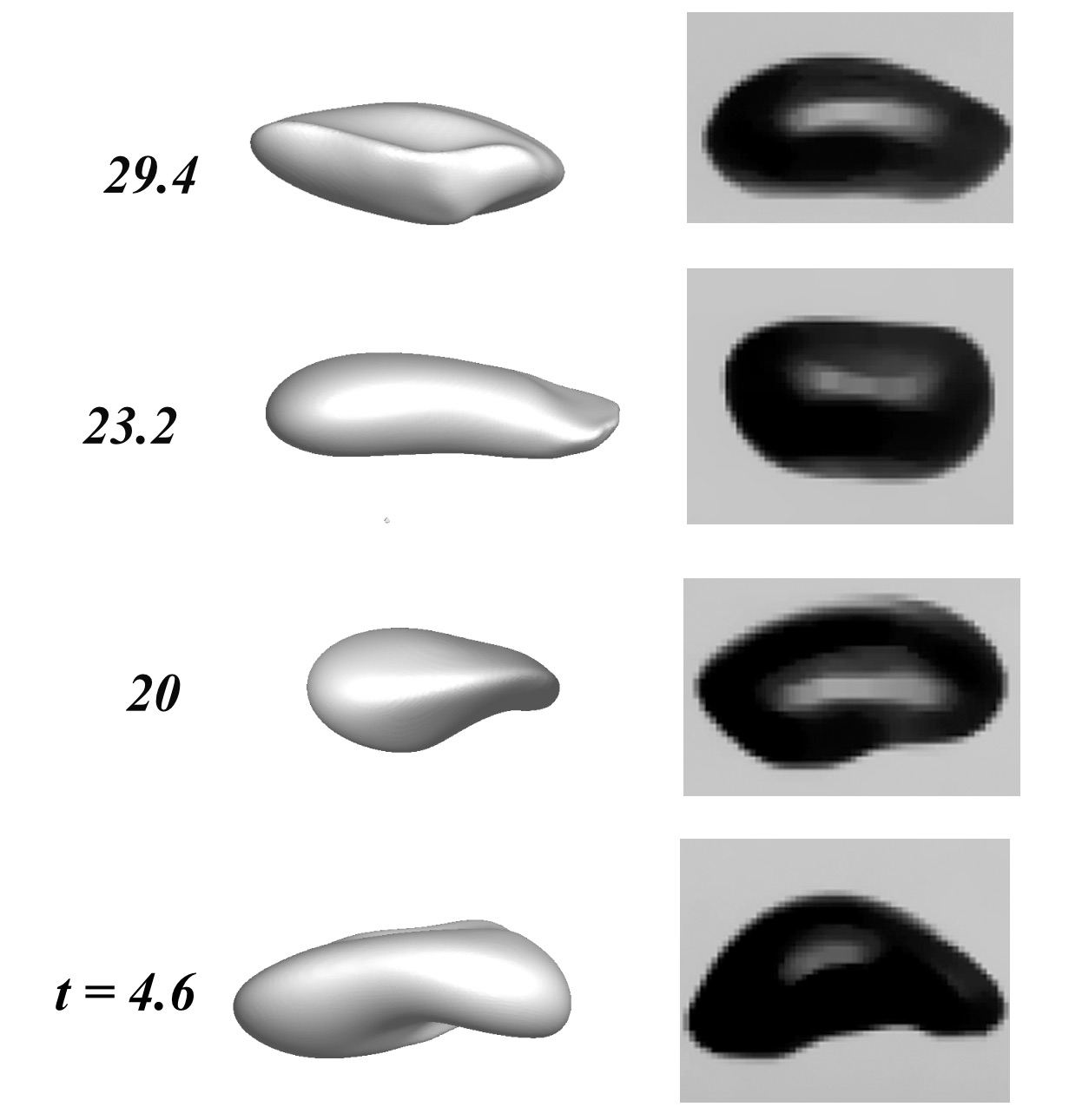} 
\caption{Evolution of paths and shapes of two region III bubbles. (a) The left panel shows zoomed view of the shapes of the bubbles at different times and the right panel shows the trajectory of the bubble (multimedia view). The initial bubble radius is 2 $mm$ and the surrounding fluid is $25\%$ of glycerol in water ($\mu_r=1.43\times10^{-3} $, $\rho_r=9.43\times10^{-4} $, $Ga = 44.06 $ and $Eo = 0.63$). (b) Evolution of bubble shapes at four time instances for an air bubble of initial radius 4.57 $mm$ and a surrounding fluid of $10\%$ of glycerol in water ($\mu_r=2.33\times10^{-3} $, $\rho_r=9.78\times10^{-4}$, $Ga=230.8$ and $Eo=3$). The left and right panels show numerical and experimental results, respectively.
}
\label{oscillatory2}
\end{figure}

Region III bubbles show extensive unsteady behaviour in the paths as well as bubble shapes. A trajectory of an air bubble ($R \approx  2 mm$) rising in $25\%$ of glycerol in water solution obtained from the experiment is shown in Fig. \ref{oscillatory2}(a). The shapes at different time instants (shown in zoomed view in the left panel) are overlapped to show the trajectory (right panel of Fig. \ref{oscillatory2}(a)). In this particular view, the trajectory shows a total deviation of six times the bubble radius about the axis of symmetry (vertical dashed line). The path of the bubble in this case is found to be spiralling (Multimedia view). In a recent paper, Cano-Lozano {\it et al.} \cite{cano2016} demonstrate that the unsteady shape deformations are related to the rotation of the bubble along the zigzag path. {It has been well known that the path oscillation of a region III bubble can be a result of either shape asymmetries, or unsteady vortex shedding, or both. However, recently, Vries \cite{antoine} found a regime of path instability where no vortex shedding was expected. Thus, without making an argument about the cause and effect of path instability, we only say that there is an intimate connection between loss of symmetry and loss of a straight trajectory. This intuition is due to the fact that any asymmetrical deformation of the bubble in the plane perpendicular to gravity would result in an imbalance of planar forces, which in turn would drive the bubble away from the axis of the domain, and vice versa.}

\begin{figure}
\centering
\hspace{0.5cm} (a) \hspace{6.0cm} (b) \\
 \includegraphics[width=0.5\textwidth]{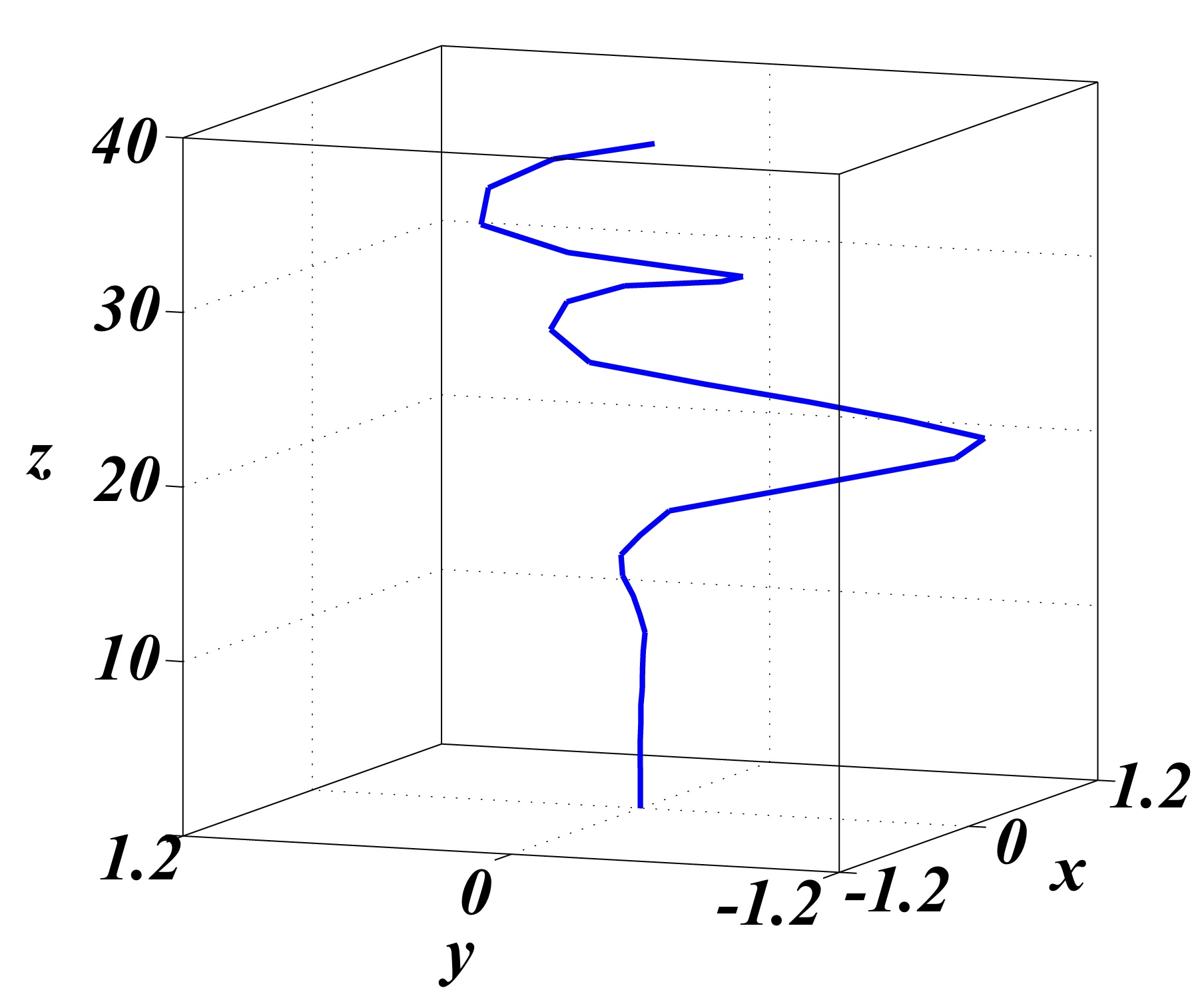} \hspace{1cm} 
\includegraphics[width=0.4\textwidth]{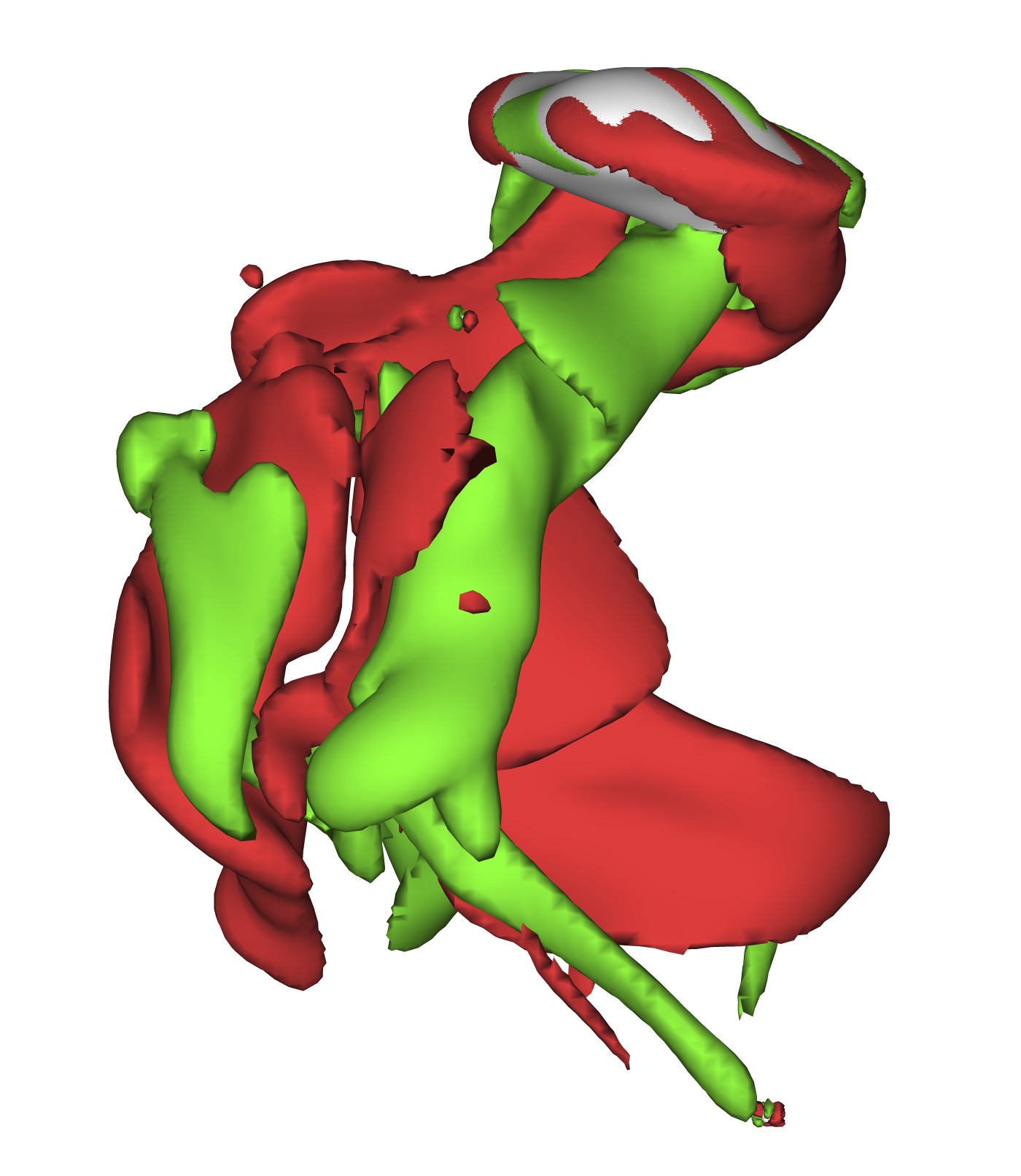} \\
\caption {(a) Trajectory and (b) iso-surfaces of the vorticity component in the $z$ direction (magnitude $\pm 0.8$) at time $t=25$ (obtained from numerical simulation). The parameters are the same as those used in Fig. \ref{oscillatory2}(b).}
\label{oscillatory3}
\end{figure}

Fig. \ref{oscillatory2}(b) shows the temporal evolution of the shape of another region III bubble of size, $R \approx 4.57$ $mm$, obtained from the experiment (right panel) and the numerical simulation (left panel). Note that this bubble (with $Ga=230.8$ and $Eo=3$) lies in region IV of Tripathi {\it et al.} \cite{tripathiNcomms2015}) (satellite breakup). This bubble falls into region IV based on their numerical simulations with $\rho_r = 10^{-3}$ and $\mu_r = 10^{-2}$. The experimental results for this case are obtained for an air bubble rising in a solution of 10\% glycerol in water with $\rho_r=9.78\times10^{-4}$ and $\mu_r=2.33\times10^{-3} $. In the numerical simulation, when we consider the modified viscosity and density ratios, i.e. $\mu_r=2.33\times10^{-3} $, $\rho_r=9.78\times10^{-4}$ instead of $\mu_r=10^{-2} $, $\rho_r=10^{-3}$ as considered by Tripathi {\it et al.} \cite{tripathiNcomms2015}, it is observed that the bubble now does not break and behaves like a region III bubble, which agrees with experiment qualitatively. These results also possibly explain the discrepancy observed in the boundary separating region III and breakup (regions IV and V) (i.e., $Ga>110$ and $1<Eo<10$) between our experimentally obtained phase diagram (Fig. \ref{fig1}) with that of Tripathi {\it et al.} \cite{tripathiNcomms2015}.

Numerical simulation shows that this bubble follows a wobbling motion as shown in Fig. \ref{oscillatory3}(a). It is observed that although the shape deformation happens at a very early time $(t=4.6)$ (see Fig. \ref{oscillatory2}(b)), the bubble travels in a straight path till $t \approx 15$. This is in accordance with the finding of Cano-Lozano {\it et al.} \cite{cano2016}, as discussed above. The iso-surfaces of the vorticity component in the $z$ direction at $t = 25$ are shown in Fig.  \ref{oscillatory3}(b). The pair of streamwise vortices result in a lift force, which is responsible for the nonzero horizontal velocity component of the bubble. It is also found that the occurrence of streamwise vorticity in the wake coincides with the path instability.


\subsubsection{Effect of viscosity and density ratios}

\begin{figure}
\centering
(a) \hspace{2.0cm} (b) \hspace{2.0cm} (c) \hspace{2.0cm} (d) \hspace{1.5cm} \\
 \includegraphics[width=0.8\textwidth]{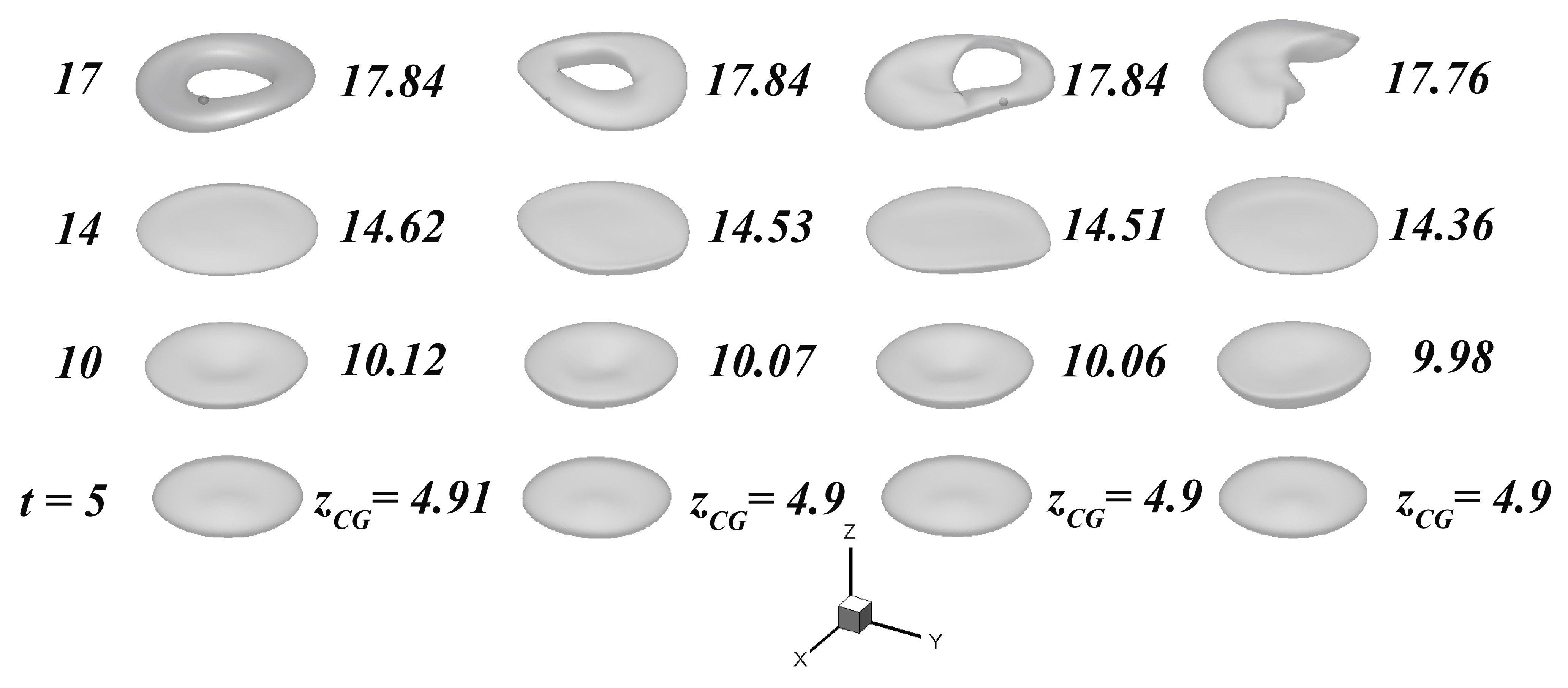} 
 \caption{Effect of viscosity ratio on the evolution of shapes: (a) $\mu_r=10^{-1}$,  (b) $\mu_r=10^{-2}$,  (c) $\mu_r=10^{-3}$ and (d) $\mu_r=10^{-4}$. The rest of the parameters are $\rho_r=10^{-3}$, $Ga=230.8$ and $Eo=3$. The locations of center of gravity, $z_{CG}$ of the bubble at different times are also shown. The shapes shown in panel (b) correspond to a region V bubble (central breakup) in Tripathi {\it et al.} \cite{tripathiNcomms2015}. 
}
\label{effectmur}
\end{figure}

\begin{figure}
\centering
(a) \hspace{2.0cm} (b) \hspace{2.0cm} (c) \hspace{2.0cm} (d) \hspace{1.5cm} \\
 \includegraphics[width=0.8\textwidth]{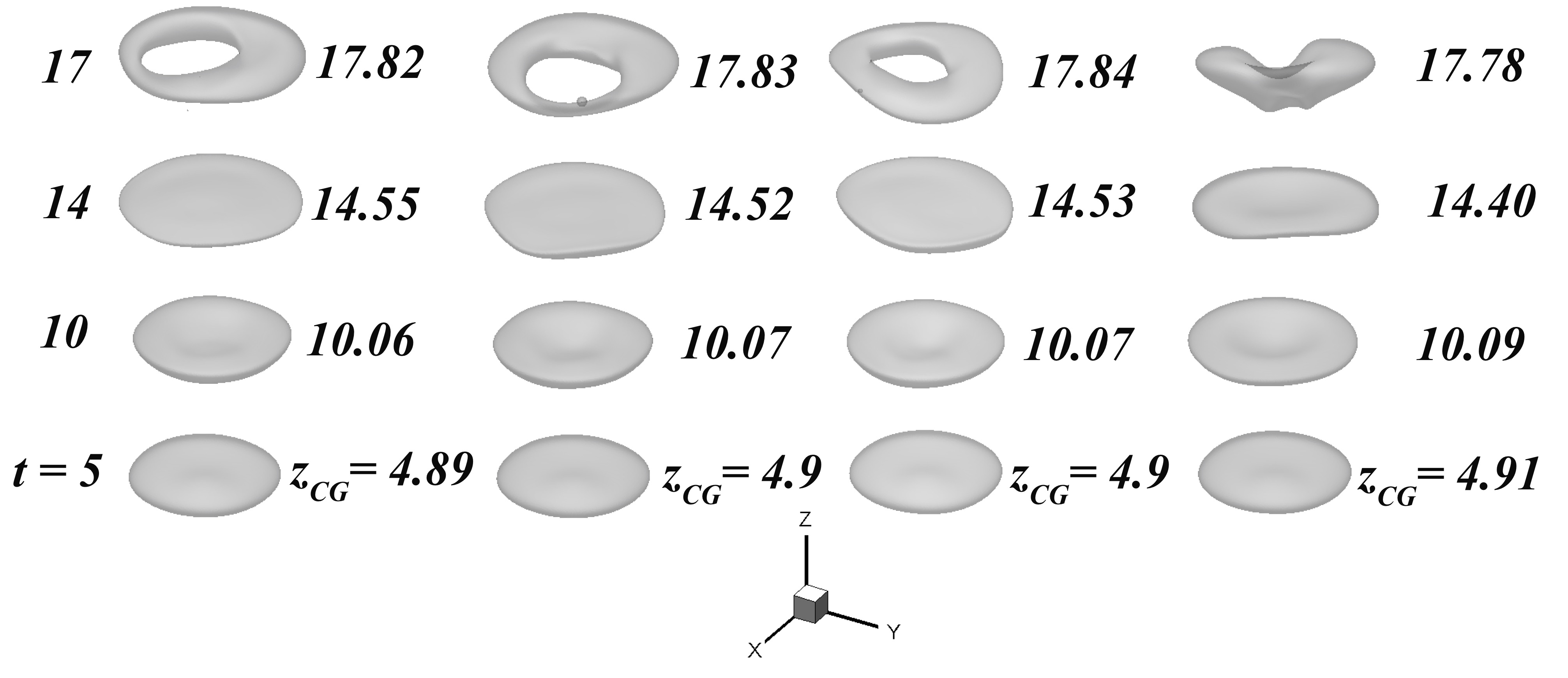} 
\caption{Effect of density ratio on the evolution of shapes: (a) $\rho_r=10^{-1}$,  (b) $\rho_r=2 \times 10^{-3}$,  (c) $\rho_r=10^{-3}$ and (d) $\rho_r=10^{-4}$. The rest of the parameters are $\mu_r=10^{-2}$, $Ga=230.8$ and $Eo=3$. The locations of center of gravity, $z_{CG}$ of the bubble at different times are also shown. The shapes shown in panel (c) correspond to a region V bubble (central breakup) in Tripathi {\it et al.} \cite{tripathiNcomms2015}. 
}
\label{effectrhor}
\end{figure}

\begin{figure}
\centering
\hspace{0.5cm} (a) \hspace{6.0cm} (b) \\
 \includegraphics[width=0.4\textwidth]{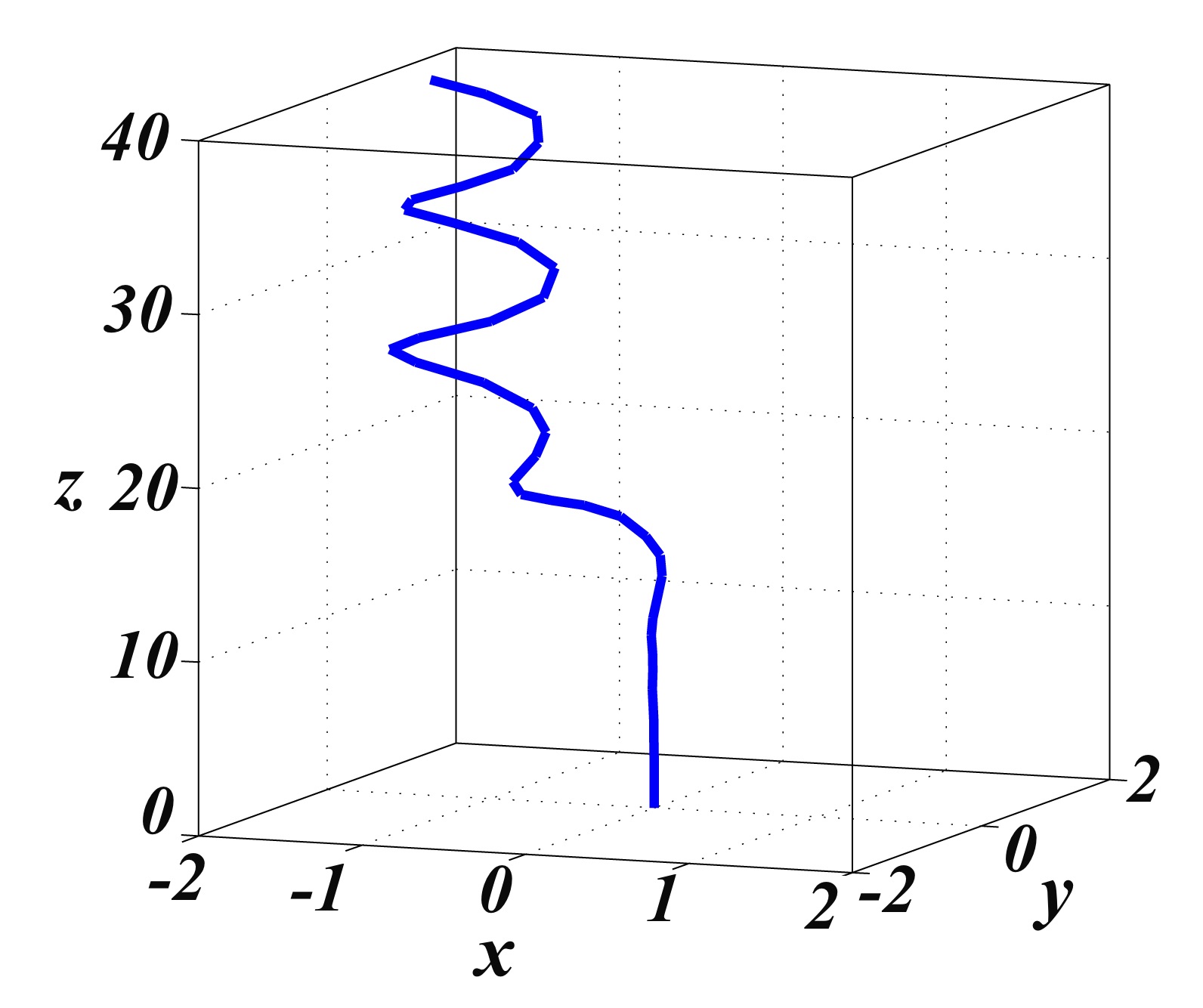} \hspace{1cm} 
\includegraphics[width=0.4\textwidth]{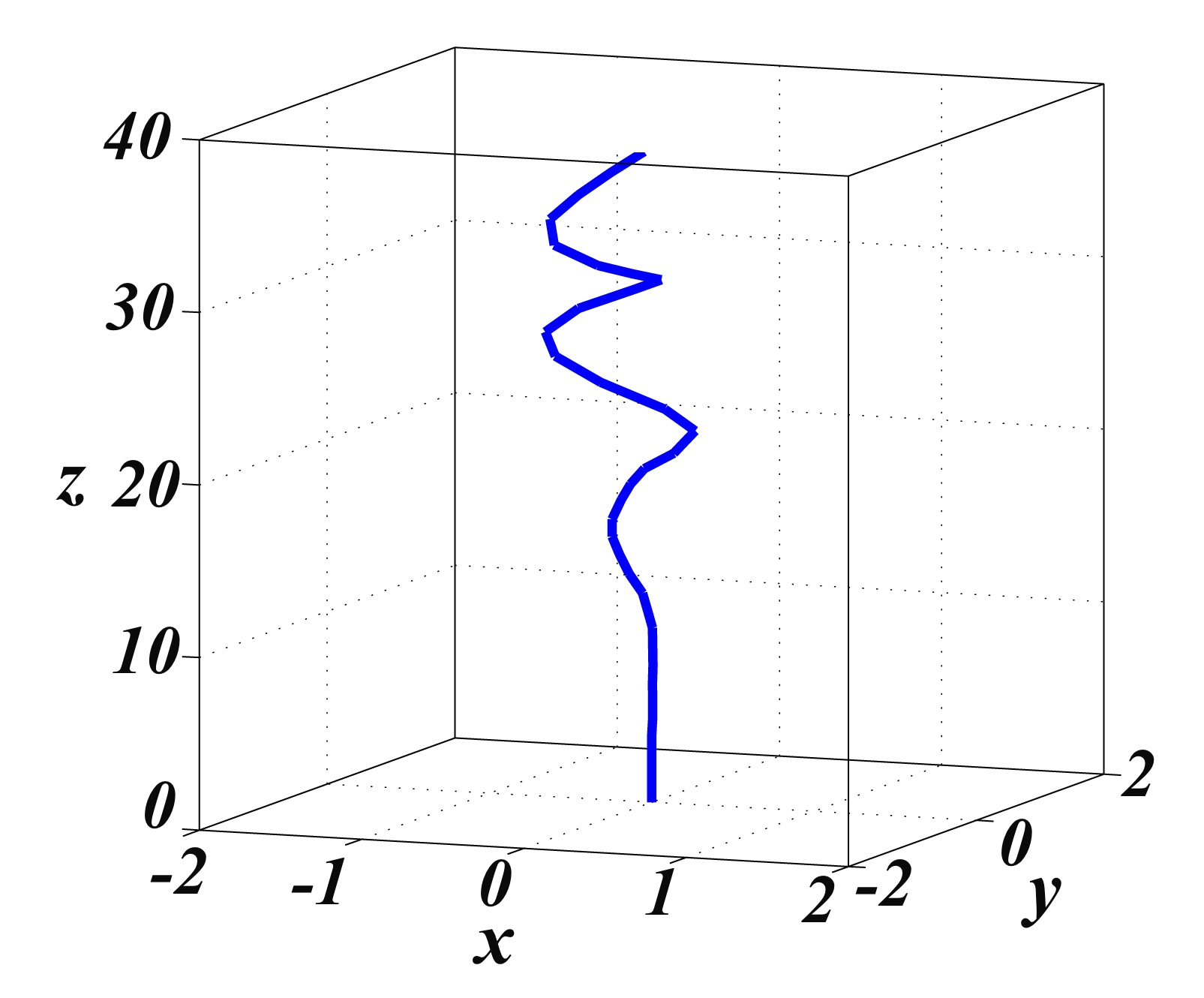} \\
\caption {Trajectories of the bubble for (a)  $\mu_r=10^{-4}$, $\rho_r=10^{-3}$, $Ga=230.8$, $Eo=3$, and (b) $\mu_r=10^{-2}$, $\rho_r=10^{-4}$, $Ga=230.8$, $Eo=3$.}
\label{effectmur_trj}
\end{figure}

The result presented above (Fig. \ref{oscillatory2}(b)) also reveals that a change in viscosity ratio (with a negligible change in density) can change the bubble region from central breakup (region V) to oscillatory region (region III). This motivated us to investigate the effect of changing the viscosity and density ratios on bubble behaviour. Numerical simulations were conducted by separately varying $\mu_r$ and $\rho_r$ while keeping the rest of the parameters constant. However, it is to be noted that the parameters considered in this section may not be feasible in experiments.

In Fig. \ref{effectmur}, $\mu_r$ is varied from $10^{-1}$ to $10^{-4}$ while the rest of the parameters are kept fixed at $\rho_r=10^{-3}$, $Ga=230.8$ and $Eo=3$. It can be seen that for $\mu_r \ge 10^{-3}$ (Fig. \ref{effectmur}(a), (b) and (c)), the bubble undergoes central breakup to form a doughnut-like or toroidal shape, which becomes unstable at later times and breaks down into smaller bubbles (not shown). It is found that bubbles in Fig. \ref{effectmur}(a), (b) and (c) travel in a straight path. This is consistent with the finding of  Tripathi {\it et al.} \cite{tripathiNcomms2015} for an air-water system ($\mu_r=10^{-2}$ and $\rho_r=10^{-3}$) for $Ga=230.8$ and $Eo=3$. Decreasing the viscosity further to $\mu_r=10^{-4}$, the bubble behaves like a region III (oscillatory/wobbling) bubble (see Fig. \ref{effectmur}(d)). The trajectory of the bubble for $\mu_r=10^{-4}$ is shown in Fig. \ref{effectmur_trj}(a). We found that vortex shedding behind the bubble (which occurs for $\mu_r=10^{-4}$ for this set of parameters) promotes this oscillatory motion by adjusting its shape without allowing it to break. 

The effect of density ratio (varying from $10^{-1}$ to $10^{-4}$) on bubble shapes for $\mu_r=10^{-2}$, $Ga=230.8$ and $Eo=3$ is presented in Fig. \ref{effectrhor}. It can be seen that for $\rho_r \ge 10^{-3}$ (Fig. \ref{effectrhor}(a), (b), (c)) the bubble undergoes topological change to form a doughnut-like shape (region V of Tripathi {\it et al.} \cite{tripathiNcomms2015}), but behaves like an oscillatory bubble (region III) for $\rho_r = 10^{-4}$ (see Fig. \ref{effectrhor}(d)). The oscillatory path of this bubble is shown in Fig. \ref{effectmur_trj}(b). On a separate note, it is mentioned here that a liquid drop falling in air ($\mu_r=57$ and $\rho_r=1000$) never does wobbling motion \cite{Agrawal17}. 

The mechanisms behind the central breakup observed for large liquid-to-gas density and viscosity ratios are discussed below. In the literature, two types of mechanisms were suggested for the toroidal breakups: (i) inertial upward jet mechanism \cite{walters} and (ii) downward jet pinch-off mechanism \cite{collins,bonometti06a}. In the inertia dominated regime, as the bubble rises, it deforms by the additional pressure generated due to the hydrostatic pressure head equivalent to $2 \rho_o  g R$ between the top and bottom poles of the bubble. This in turn creates an upward liquid jet which squeezes the bubble in the vertical direction. This deformation is counteracted by the surface tension force, in general. In this competition between the inertia and surface tension forces, if inertia wins, then the bubble breaks from the centre to form a toroidal bubble. {The downward jet pinch-off could be explained as follows. The rise in pressure at the front pole of the bubble is balanced by the capillary pressure and the normal viscous stress. However, if the capillary effects are small (high $Eo$) and if the viscous forces are also small as compared to the inertial forces (high $Ga$), the pressure excess at the front pole causes a downward jet to destabilize the interface \cite{bonometti06a}.}


We observe a third kind of toroidal breakup which occurs for a much higher value of surface tension although it resembles the downward jet mechanism. It can be seen in Figs. \ref{effectmur} and \ref{effectrhor} that the bubble attains a disc-like shape and keeps on expanding its circumference, eventually resulting in a breakup from the center. It is to be noted that a large fraction of air accumulates in the peripheral regions rather than the central part, thus making the bubble increasingly thinner at the center. This could be explained as follows. The formation of counter-rotating vortices in the wake, move the fluid outwards from the center of the bubble. This action causes the bubble to form an increasingly thinner core which punctures at a later time. Moreover, the bubble remains flat in these cases due to high inertia.

\subsection{Breaking-up bubbles: bubbles in region IV}
\label{sec:region2}

\begin{figure}
\centering
\hspace{0.5cm} (a) \hspace{1.5cm} (b) \\
\includegraphics[width=0.4\textwidth]{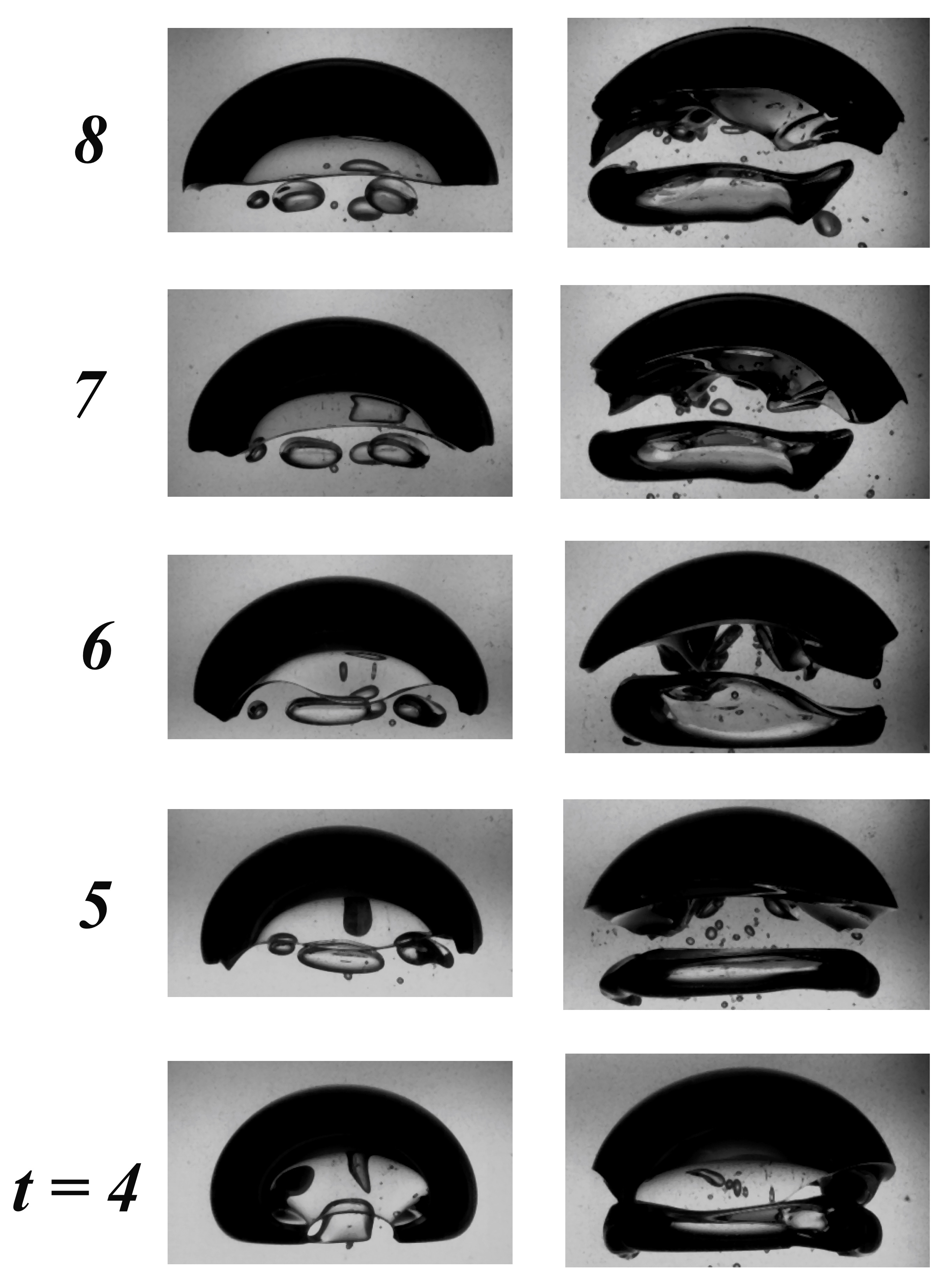} 
\caption{Evolution of breakup bubbles (region IV): (a) $70 \%$ of glycerol in water and 22.8 $mm$ bubble radius ($\mu_r=1.73\times10^{-4} $, $\rho_r=8.46\times10^{-4} $, $Ga = 221.5$ and $Eo = 92.1$), and (b) $80 \%$ of glycerol in water and 26.7 $mm$ bubble radius ($\mu_r=5.9\times10^{-5} $, $\rho_r=8.27\times10^{-4} $, $Ga = 171$, $Eo = 131$). }
\label{breakup}
\end{figure}

The numerical phase diagram of Tripathi {\it et al.} \cite{tripathiNcomms2015} indicates that there are two types of breakups for bubbles, namely peripheral (region IV) and central (region V) breakups. However, in the experimental phase diagram (Fig. \ref{fig1}) the central breakup is not observed. Within the region classified as breakup, we only observed peripheral-type breakups, which we have further classified as skirted, satellite and toroidal breakups. In the skirted breakup region, a thin skirt forms on the periphery of the bubble which breaks to form small satellite bubbles rising in the wake region following the main bubble. The satellite and toroidal breakups seem to happen arbitrarily and there is no specific boundary between them. Two recorded cases of each of these breakups are shown in Fig. \ref{breakup}(a) and (b), respectively. In case of the satellite breakup (Fig. \ref{breakup}(a)), the bubble breaks near the center to form satellite bubbles (at the wake) and a spherical cap bubble. For toroidal breakup (Fig. \ref{breakup}(b)), a ring-like structure is detached from the main body of the bubble. Walters \& Davidson \cite{walters} have predicted this kind of breakup for an initially spherical bubble rising in an inviscid liquid. They found the toroidal bubble to be stable, but we see that the presence of unsteadiness in the wake of the bubble makes the system of bubbles unstable. 

\section{Concluding remarks}
\label{sec:conc}
The dynamics of a rising air bubble inside aqueous solutions of glycerol is investigated resulting in a phase plot in the Galilei and E\"{o}tv\"{o}s numbers plane, which separates four distinct regions in terms of bubble behaviour, namely axisymmetric, skirted, spiralling and break-up regions. The experimental results are compared with those of numerical simulations to show the similarities and differences. The differences observed in the breakup region are attributed to the difficulty in creating a perfectly spherical bubble in experiment, particularly when the size of the bubble is large, i.e. for high $Ga$ and high $Eo$. Apart from Reynolds and  E\"{o}tv\"{o}s numbers, which were thought to be the only important governing parameters for rising bubbles in air-liquid systems, our results show that the actual density and viscosity ratios are also required to describe the bubble dynamics accurately, especially in the parameter space close to the region boundaries in the phase-plot.
\\
\\
\noindent {\bf Acknowledgement:} We also like to thank Ms. Susree Modepalli of IIT Hyderabad for her help during the viscosity measurements. The authors also thank Prof. Sumohana Channappayya of IIT Hyderabad for his help in post-processing the images obtained from our experiments. K. C. S. thanks Indian National Science Academy for their financial support. B. K. thanks the Department of Science and Technology, India.


\end{document}